\documentclass[%
 reprint,
 amsmath,amssymb,
prapplied,
]{revtex4-2}

\usepackage{graphicx}
\usepackage{dcolumn}
\usepackage{bm}
\usepackage{comment}
\usepackage{amsmath}
\usepackage[colorlinks=true,bookmarks=false,citecolor=blue,urlcolor=blue]{hyperref} 
\usepackage{upgreek}

\newcommand{\ind}{\hspace{1 em}}
\newcommand{\textss}[1]{\scriptsize \mbox{#1}}

\DeclareMathAccent{\wtilde}{\mathord}{largesymbols}{"65}
\newcommand{\Lnl}{ L_{\textss{NL}}^{(2)} }
\newcommand{\Lgvm}{L_{\textss{GVM}}}

\begin{document}

\preprint{APS/123-QED}

\title{Broadband ultraviolet-visible frequency combs from cascaded high-harmonic generation in quasi-phase-matched waveguides}

\author{$^1$Jay Rutledge}
\author{$^1$Anthony Catanese}
\author{$^{2,3}$Daniel D. Hickstein} 
\author{$^{2,3}$Scott A. Diddams}
\email{scott.diddams@nist.gov}
\author{$^1$Thomas K. Allison}
\email{thomas.allison@stonybrook.edu}
\author{$^{2,3}$Abijith S. Kowligy}

\affiliation{$^1$Stony Brook University, Stony Brook, New York 11794-3400, USA}
\affiliation{$^{2}$National Institute of Standards and Technology, Boulder, Colorado 80305, USA}
\affiliation{$^{3}$Department of Physics, University of Colorado Boulder, Boulder, Colorado 80309, USA}

%



\date{\today}

\begin{abstract}


High-harmonic generation (HHG) provides short-wavelength light that is useful for precision spectroscopy and probing ultrafast dynamics. We report efficient, phase-coherent harmonic generation up to 9th-order (333 nm) in chirped periodically poled lithium niobate waveguides driven by phase-stable $\leq$12-nJ, 100 fs pulses at 3 $\upmu$m with 100 MHz repetition rate. A mid-infrared to ultraviolet-visible conversion efficiency as high as 10\% is observed, amongst an overall 23\% conversion of the fundamental to all harmonics. We verify the coherence of the harmonic frequency combs despite the complex highly nonlinear process. Numerical simulations based on a single broadband envelope equation with quadratic nonlinearity give estimates for the conversion efficiency within approximately 1 order of magnitude over a wide range of experimental parameters. From this comparison we identify a dimensionless parameter capturing the competition between three-wave mixing and group-velocity walk-off of the harmonics that governs the cascaded HHG physics. These results can inform cascaded HHG in a range of different platforms.

\end{abstract}

\maketitle

\section{\label{sec:level1}Introduction}

When intense ultrashort-pulse lasers interact with matter, high-order harmonics of the fundamental laser frequency can be generated. The phenomena of high-harmonic generation (HHG) is observed in all forms of matter from atomic and molecular gasses \cite{Ferray_JPhysB1988, McPherson_JOSAB1987}, to solids \cite{Ghimire_NatPhys2011}, liquids \cite{DiChiara_OptExp2009, Luu_NatComm2018}, and dense plasmas \cite{Dromey_NatPhys2006}. Understanding the physical mechanisms behind HHG, particularly harmonics generated in the regime where the light-matter interaction is not well described by perturbation theory, has been a subject of intense research for the past three decades. Furthermore, the harmonic light is now used for a variety of applications ranging from precision frequency metrology \cite{Kandula_PRL2010, Cingoz_Nature2012} to femtosecond spectroscopy \cite{Allison_OptLett2010, Corder_StructDyn2018} and microscopy \cite{Sandberg_PRL2007, Mikkelsen_RSI2009} to attosecond spectroscopy \cite{Chini_NatPhot2014_2, Krausz_RMP2009}. 

\ind While HHG in the non-perturbative regime can provide very high-order harmonics, even extending into the extreme ultraviolet and soft x-ray \cite{Schnurer_PRL1998, Popmintchev_Science2012}, the conversion efficiency is inherently very low compared to more conventional nonlinear optical phenomena based on $\chi^{(n)}$ processes \cite{Boyd:2003}. Furthermore, non-perturbative nonlinear light-matter interaction is usually only accessed with ultrashort pulses with peak powers $>$ 100 MW, precluding applications in chip-scale photonic devices or high repetition-rate frequency combs. Perturbative light-matter interactions can potentially offer an alternative route to HHG via the cascading of many lower-order processes. As is well known for other nonlinear optical phenomena \cite{DeSalvo_OptLett1992, Liu_OptLett1999}, under the right conditions the effective strength of cascaded nonlinear processes can exceed direct processes by orders of magnitude. For example, recently Couch et al. \cite{Couch_Optica2020} reported efficient high-harmonic generation to the vacuum ultraviolet via cascaded $\chi^{(3)}$ processes in Xe-filled waveguides driven with 10 $\upmu$J pulses. 

\ind For applications requiring a high repetition rate, such as frequency comb generation, pulse energy thresholds in the nJ range or lower are desired. High-order harmonic generation in solids generally has peak intensity requirements much lower than gasses \cite{Ghimire_NatPhys2011, Ghimire_NatPhys2019}, and solid-state HHG has been observed in a range of materials including atomically-thin semiconductors \cite{Liu_NatPhys2017}, metallic nanostructures \cite{Liu_NatPhys2018}, and fractal-poled crystals \cite{Park_OptLett2017}. Previously, Hickstein et al. \cite{Hickstein_Optica2017, Hickstein_CLEO2018} reported the observation of high harmonics, up to the 13th order, from periodically-poled lithium niobate (PPLN) waveguides driven with 4.1 $\upmu$m, 200 fs pulses with only $\sim$10 nJ of energy. The harmonics covered the entire visible spectrum into the ultraviolet with remarkable conversion efficiencies in the percent range. This low pulse energy threshold can enable broadband frequency conversion at high repetition rate, which is essential for frequency comb applications. However, the previous report from Hickstein et al. left important questions regarding the physics of the PPLN HHG process in these waveguides unanswered, such as whether the HHG process was due to perturbative or non-perturbative nonlinear optics. The phase-coherence of the generated harmonics also was not determined. Coherence of the short-wavelength waveguide emissions is not \emph{a priori} obvious for such highly nonlinear processes, as has been studied extensively for supercontinuum generation with high soliton number \cite{Dudley_2006}.

\begin{figure*}[t]
\centering
\includegraphics[width=\linewidth]{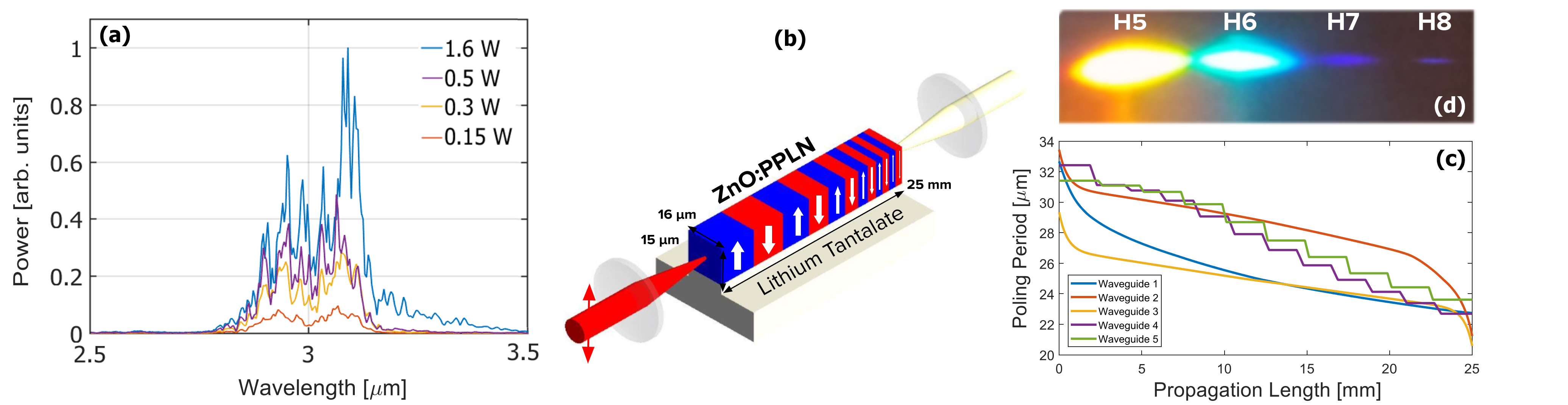}
\caption{Schematic of the experimental setup. \textbf{(a)} Mid-IR pump spectrum exiting the waveguide at different incident powers. The lower power spectrum is similar to the incident spectrum. \textbf{(b)} Waveguide and coupling apparatus consisting of ZnO:PPLN waveguides on a lithium tantalate substrate. Here the poling period is greatly exaggerated for illustrative purpose. \textbf{(c)} Poling profiles of the five waveguides. 
\textbf{(d)} Harmonic light dispersed with a prism and viewed on white paper.}
\label{fig:setup}
\end{figure*}

\ind In this article, we present a comprehensive study of HHG in PPLN waveguides. First, we report new measurements of PPLN HHG using phase-stable 3 $\upmu$m pulses at 100 MHz repetition rate from a recently developed mid-IR frequency comb at Stony Brook University \cite{Catanese_OptLett2020}. High-brightness HHG frequency combs are observed emerging from the waveguide covering the visible and UV up to the lithium niobate band gap, with generated powers exceeding 100 mW per harmonic. Via heterodyne detection, we explicitly verify the frequency comb structure is present in the high harmonics despite the highly nonlinear HHG process. Second, we perform detailed simulations of pulse propagation in the PPLN waveguides with $\chi^{(2)}$ nonlinearity using the nonlinear analytic envelope equation (NAEE) formalism of Conforti et al. \cite{Conforti_PRA2010, Conforti_IEEE2010}. Third, we compare both the new measurements and the simulations to the previous measurements of Hickstein et al. \cite{Hickstein_Optica2017}.

\ind In addition to our technical results of high-power broadband frequency comb generation, from this comprehensive approach we report several more general results. First, semi-quantitative agreement between the experiments and simulations including only $\chi^{(2)}$ processes, over a wide parameter range, strongly indicates cascaded processes as the dominant HHG mechanism, largely resolving the question posed by Hickstein et al. \cite{Hickstein_Optica2017}. Second, although the harmonics are generated from cascaded $\chi^{(2)}$ processes, from comparison of harmonics driven with 4.1 $\upmu$m and 3 $\upmu$m it emerges that the absolute optical frequency of the harmonics appears more important for determining the conversion efficiency than the harmonic order. This sort of ``plateau" behavior is more typically associated with non-perturbative harmonic generation. Third, although quasi-phase matching at various locations in the waveguide is important for generating the harmonics, very similar harmonic spectra are observed from waveguides with quite different poling period profiles. This quasi-universal behavior implies that although the details of phase matching in the waveguide are important, more important for cascaded waveguide HHG is the competition between three-wave mixing and group-velocity walk-off of the harmonics. Via overlaying simulation and experiment, we identify a dimensionless parameter that captures this competition, determines the onset of cascaded HHG, and provides an intuitive understanding of the key parameters governing the HHG physics. These critical understandings can inform the design of future cascaded HHG sources.

\section{3 $\upmu$\MakeLowercase{m} Experiment}
\subsection{Light source, waveguides, and detection}

Using the idler output of a high-power optical parametric amplifier previously described in \cite{Catanese_OptLett2020}, we deliver phase-stable, 100 fs pulses at 3 $\upmu$m, with 100 MHz repetition rate and up to 20 nJ of pulse energy, to the waveguide setup depicted in figure \ref{fig:setup}b. The mid-IR spectrum exiting the waveguide at low and high power is shown in figure \ref{fig:setup}a. The 3 $\upmu$m pump power at the waveguide is varied using a zero-order half wave plate (Edmund Optics) and a Brewster angle reflection from an uncoated Si substrate as a broadband polarizer. The mid-IR driver, with photon energy well below the material band gap of 4.0 eV, enables high peak intensities in the waveguide of $3 \times 10^{11}$ W/cm$^2$ at 100 MHz repetition rate without damage. On the other hand, near-IR driving with the OPA's concomitant signal beam at 1.6 $\upmu$m results in damage at power levels below the onset of significant HHG. 

\ind The waveguides (Fig. \ref{fig:setup}b) are made of lithium niobate with 7.5\% ZnO doping \cite{Castill-Torres_OptComm2013} and supported on a lithium tantalate substrate. The input and output facets are AR coated for 2800-5000 nm. The doping raises both lithium niobate's threshold for photorefractive damage and its band gap, enabling higher pump peak intensities in the waveguide and transparency for the 9th harmonic of 3 $\upmu$m at 333 nm. Each waveguide has the same cross-section ($15\times16$ $\upmu$m) and length ($25$ mm) but unique chirped poling period profile as shown in figure \ref{fig:setup}c. The pump polarization is along the extraordinary axis of the crystal (red vertical arrow in figure \ref{fig:setup}b). We test these five PPLN waveguides, three exactly as appearing in \cite{Hickstein_Optica2017}. Select waveguides are tested ``forward" (decreasing poling-period direction) and ``backward" (increasing poling-period direction) to expose possible QPM dependencies.

\ind We couple the 3 $\upmu$m frequency comb light into the waveguide using an AR-coated chalcogenide aspheric lens. The fundamental and harmonics exiting the waveguide are collimated using an AR-coated fused-silica lens. The lenses and the coupling setup here are improved from that of the previous work of Hickstein et al. \cite{Hickstein_Optica2017}, such that harmonics appear at lower pump pulse energies than previously reported. We measure an overall throughput of the setup shown in figure \ref{fig:setup}b of 60\% at low power, which we assume is dominated by the input coupling, such that in-waveguide pulse energies are calculated by multiplying the incident power by 0.6. For the rest of the paper, we refer to in-waveguide pulse energies with this assumption, although comparison with theory indicates that the energy effectively coupled into the fundamental mode of the waveguide is likely somewhat less.

\begin{figure*}[t]
\centering\includegraphics[width=\linewidth]{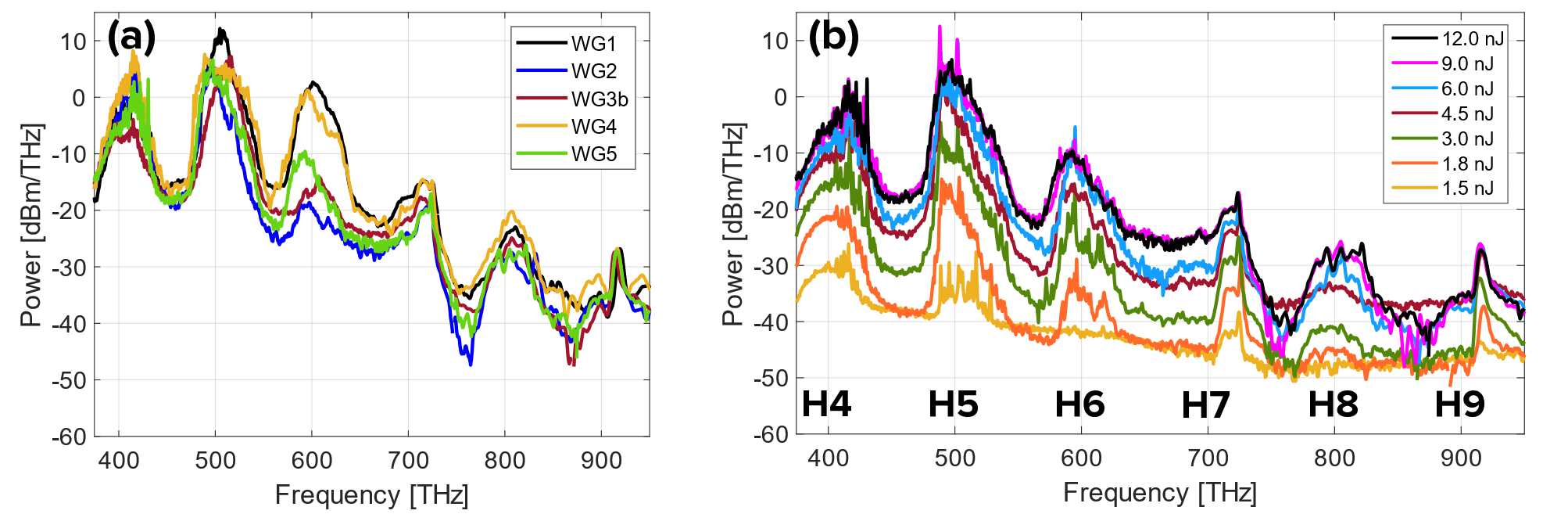}
\caption{\textbf{(a)} Comparison of measured absolute spectra for all five waveguides at 12 nJ pumping. Designation `b' for WG3 indicates that harmonics were measured in the backward (increasing poling period) pumping direction. Qualitatively similar spectra are observed from each of the waveguides. \textbf{(b)} Measured absolute spectra at various pump pulse energies in Waveguide 5. The harmonics first appear at 1.5 nJ and saturate around 9 nJ.}
\label{fig:spectra}
\end{figure*}

\ind For measuring harmonic spectra in the 300-800 nm range (H4-H9) we use a 1/3-m, Czerny-Turner monochromator with a 1200 groove/mm diffraction grating blazed for 300 nm. White light from the waveguide is focused at the entrance slit of this monochromator using an uncoated fused silica lens. Harmonics are then detected at the exit slit of the monochromator by a Si photodiode (Thorlabs model FDS1010) and a transimpedance amplifier with a typical gain $10^7$ V/A. To avoid PD saturation when coupled pump pulse energies exceed 6 nJ, we record harmonics H4-H7 separately with lower transimpedance gain. We then combine data recorded with both gain settings to produce the curves shown in figure \ref{fig:spectra}. We determined the dynamic range and resolution of the setup to be 23 dB and 0.7 nm, respectively, by measuring the response to a narrow-linewidth HeNe laser under typical system settings. 

\ind The relative wavelength-dependent response of the monochromator is determined by measuring the spectrum of an independently calibrated Xe-Hg lamp. The harmonic spectra are calibrated on an absolute scale by isolating the 6th-harmonic (500 nm) via a prism (Fig. \ref{fig:setup}d) and measuring its power independently with the Si photodiode. We correct for the harmonic power for the Fresnel reflection losses of the prism, but not those of the waveguide output facet and collimating lens, since the AR coating performance at the harmonic wavelengths is not known. We estimate this could affect the results for the measured conversion efficiencies at the 10-20\% level, on average. The raw spectra are then first corrected for the relative response determined with the Xe-Hg lamp and then scaled such that the integrated power in the 6th harmonic signal measured at the monochromator exit agrees with the power measured in the prism-isolated 6th harmonic. Once the appropriate scale factor is determined, we observe excellent agreement between the power measured via the prism-isolation scheme and integrating the spectrum over a wide range of pump powers. 

\ind \ind The integrated power in all the harmonics (including H2 and H3, which were out of range of the monochromator) is determined by inserting a filter that transmits all the harmonics but the fundamental, and measuring the resulting pink beam (H2-H9) power on a thermal power meter. We also measure the mid-IR pump spectrum using a Fourier transform spectrometer (ARCoptix FTIR-Rocket). We searched for any generated infrared light beyond 3 $\upmu$m from the waveguide, due possibly to DFG processes \cite{Kowligy_OptLett2018}, but no such light was detected. However, the output mid-IR spectrum does show signatures of significant phase-modulation of the pump pulse (Fig. \ref{fig:setup}a).

\subsection{Experimental results}

We measure harmonic emissions from all five waveguides at the highest pump pulse energy (12 nJ). Each waveguide is tested in the forward direction, except for waveguide 3 in the backward direction. 
 Remarkably, all waveguides produce similar spectra (Fig. \ref{fig:spectra}a) except in some cases a suppressed 6th-harmonic at 600 THz, suggesting that while quasi-phase-matching is indeed important \cite{Hickstein_Optica2017}, the overall HHG efficiency appears not critically dependent on the precise poling profile.

\ind We further investigated the harmonics from waveguide 5 by recording spectra at pump powers ranging over an order of magnitude (Fig. \ref{fig:spectra}b). While no harmonics are observed above the detection limit of our setup at pulse energies of 1.2 nJ and below, we observe all but H6 and H8 to appear at 1.5 nJ. The harmonic yields then increase with pumping power until they saturate around 9 nJ. The harmonic spectral lineshapes do not change significantly upon increasing pump power, despite the yields ranging over 3 orders of magnitude. The narrow spikes in the spectrum are reproducible over many scans. In all waveguide cases, the 7th and 9th harmonic appear displaced to higher energy than expected from a direct multiplication of the fundamental frequency. No harmonics are observed above the 9th harmonic, which we attribute to strong absorption at photon energies above the lithium niobate band gap at 967 THz \cite{Castill-Torres_OptComm2013}.
 
\begin{figure}[t]
\centering\includegraphics[width=\linewidth]{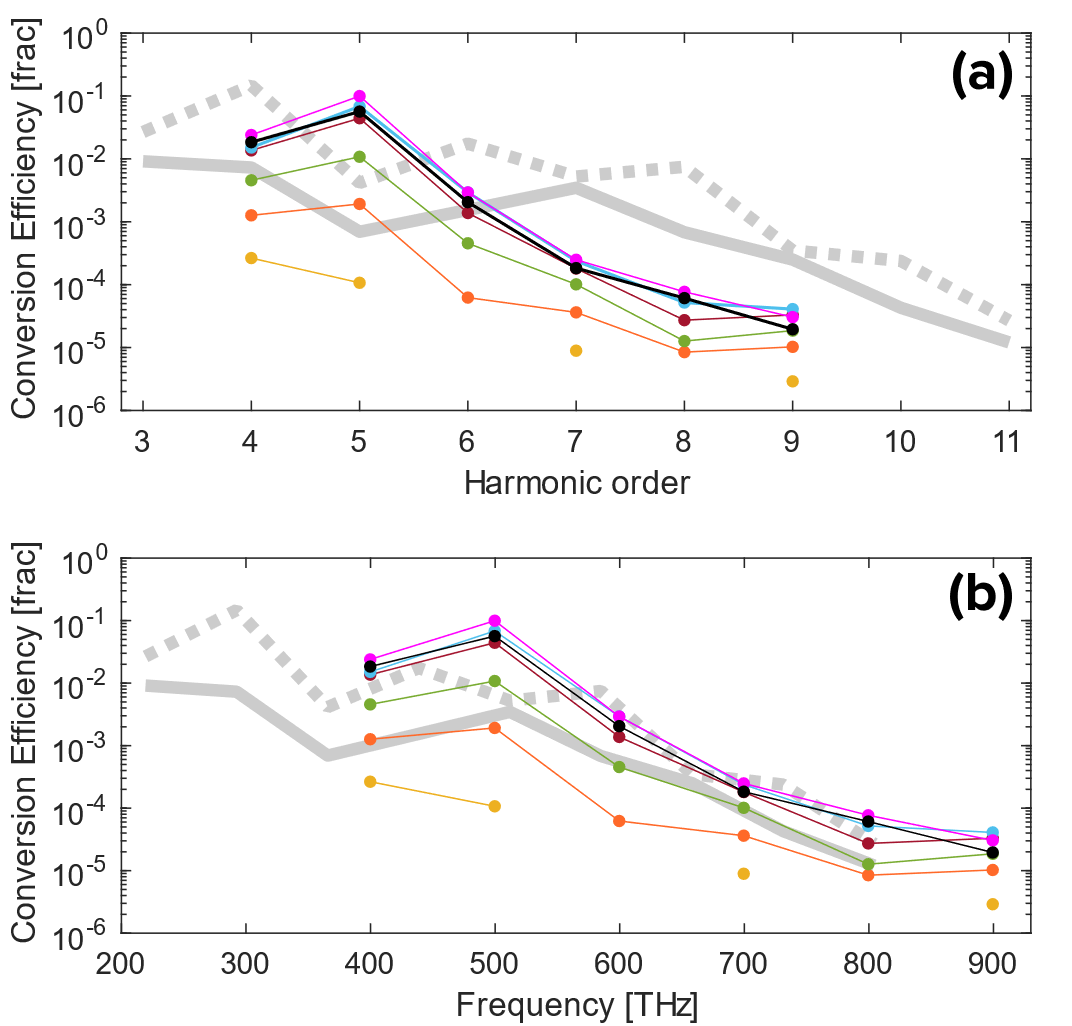}
\caption{Integrated harmonic conversion efficiencies at various pump powers corresponding to the Waveguide 5 spectra in figure \ref{fig:spectra}b. Overlaying with the 4.1 $\upmu$m experiment for both Waveguide 5, forward at 25 nJ (gray solid) and Waveguide 2, backward at 10 nJ (gray dashed) suggests that correspondence is not with  harmonic order \textbf{(a)}, but rather with absolute harmonic frequency \textbf{(b)}.}
\label{fig:CE}
\end{figure}

\ind We integrate the Waveguide 5 spectra at various powers in figure \ref{fig:spectra}b to determine the harmonic conversion efficiency, and plot the conversion efficiency versus both harmonic order (Fig. \ref{fig:CE}a) and harmonic frequency (Fig. \ref{fig:CE}b). The integrated power in harmonics 4-9, in the UV-Visible range, exceeds 10 \% of the pump power. The total power estimated in all harmonics (including 2 and 3) is measured to exceed 23\% of the pump power. Also overlaid are results from the previous work of Hickstein et al. using 4.1 $\upmu$m pulses \cite{Hickstein_Optica2017} for Waveguide 2 (gray dashed) and Waveguide 5 (gray solid). Interestingly, the experiments best coincide when compared in terms of absolute harmonic frequency as opposed to harmonic order.

\begin{figure}[t]
\centering\includegraphics[width=\linewidth]{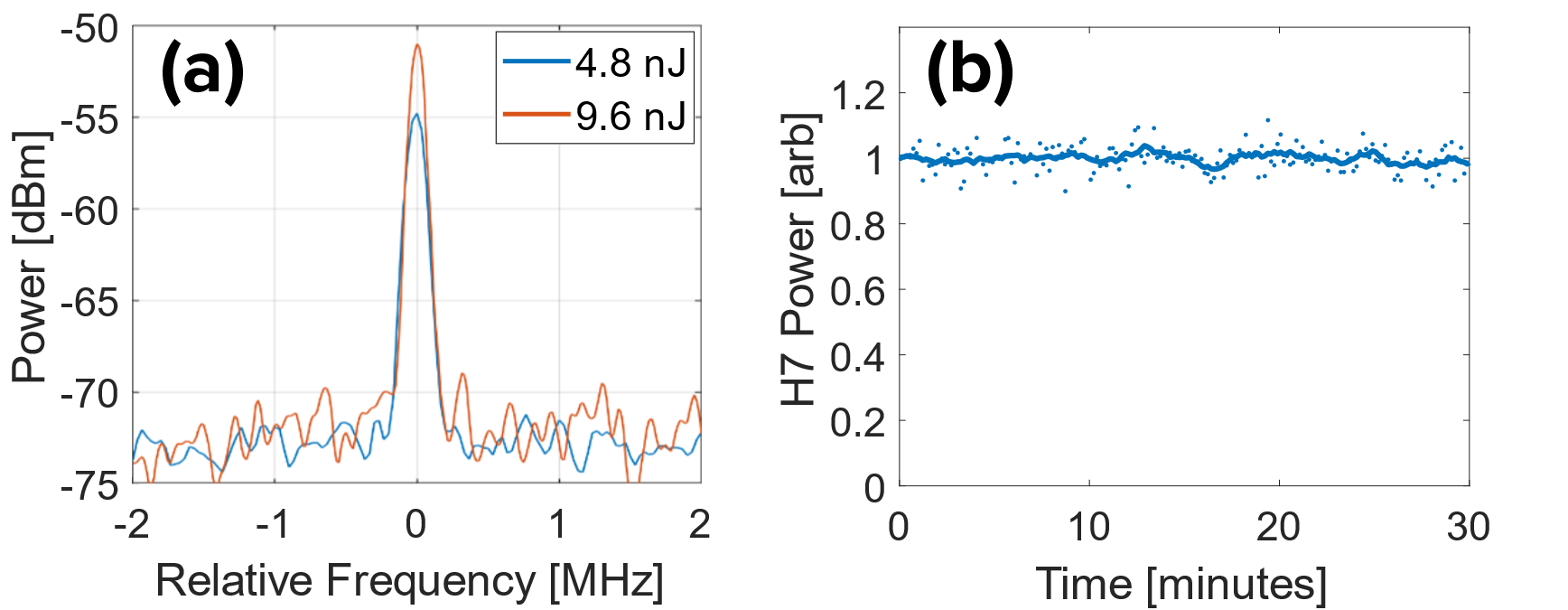}
\caption{\textbf{(b)} Heterodyne beat notes (100 kHz resolution bandwidth) between the 6th-harmonic from the waveguide and 2nd-harmonic of the Yb:fiber pump comb used in the OPA, verifying the harmonic coherence. \textbf{(b)} Measured power in a 0.7 nm bandwidth at the 7th-harmonic (428 nm) over a 30 minute duration, highlighting the long-term stability of this experimental setup. Dotted points represent 10 second measurement bins and the solid line is their smoothed average.}
\label{fig:beatstab}
\end{figure}

\begin{figure*}[t]
\centering\includegraphics[width=\linewidth]{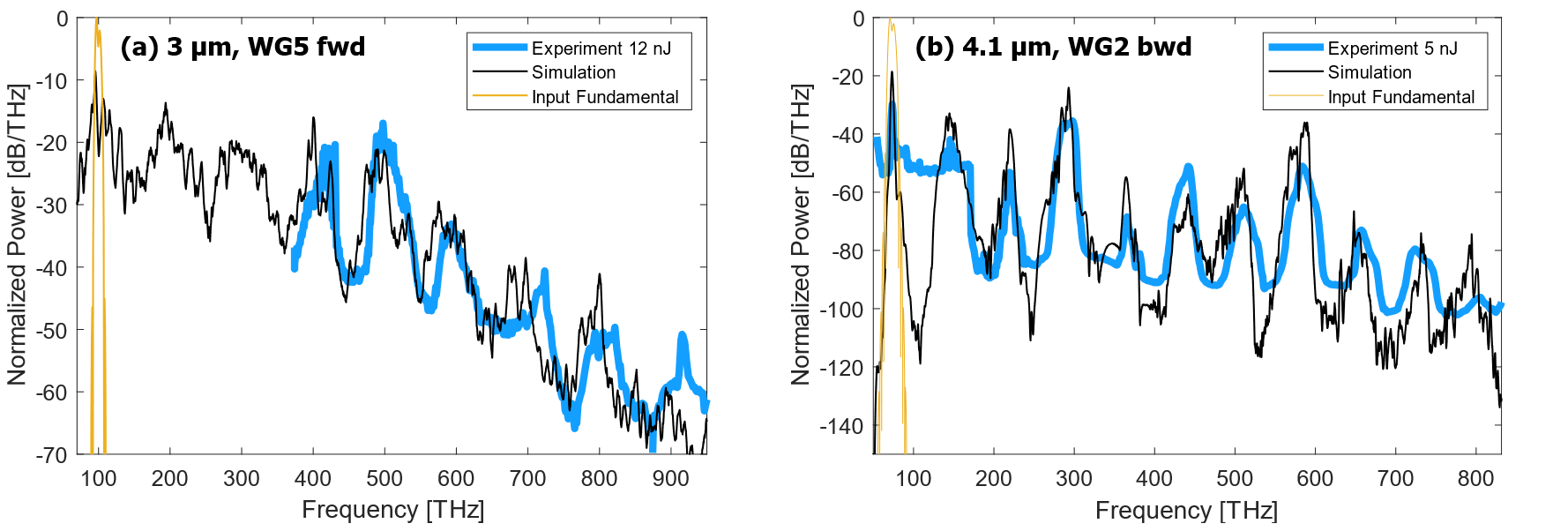}
\caption{\textbf{(a)} Overlay of simulated output spectrum (thin black line) and input pump spectrum (thin yellow line) having $\mathcal{R} = \Lgvm/\Lnl = 5.74$ with experimental data (thick blue line) for 3 $\upmu$m, 12 nJ pumping in Waveguide 5, forward. \textbf{(b)} Idem for 4.1 $\upmu$m pumping in Waveguide 2, backward with $\mathcal{R} = 1.57$ and 5 nJ experimental pulse energy. Experimental data in \textbf{(b)} is from the study first reported in \cite{Hickstein_Optica2017}.
}
\label{fig:spectraoverlay}
\end{figure*}
%
\ind We verify the coherence of the harmonic light by a heterodyne measurement between the 6th-harmonic of the waveguide and the 2nd-harmonic of the OPA's 1035 nm pump comb. The pulses are combined spatiotemporally on a Si photodiode and a beatnote is observed in the radiofrequency (RF) spectrum, shown in figure \ref{fig:beatstab}a at two different pumping powers. Since the harmonics are driven with a carrier-envelope phase stable comb with $f_0 = 0$, the harmonics should also have $f_0 = 0$. Meanwhile, the 2nd-harmonic of the OPA pump inherits the non-zero carrier-envelope frequency of the pump via $f_{0,2\omega} = 2f_{0,\textss{pump}}$ such that the beat note appears at $2f_{0,\textss{pump}}$, which we verified by changing the $f_0$ of the pump comb and observing the beat note shift appropriately. We observed this free-running beatnote to be resolution-bandwidth limited at 100 kHz in both cases. Thus, the coherence of the driving frequency comb is preserved despite the complex harmonic generation process.

\ind Figure \ref{fig:beatstab}b displays the power in a 0.7 nm bandwidth at the 7th-harmonic (428 nm) isolated with the monochromator, sampled every 2.5 ms with 100 Hz electrical bandwidth over a thirty minute duration. This shows $< 0.1$ integrated RMS intensity noise, and more importantly long-term stability despite the high average powers in the micron-scale waveguides. This demonstrates the usability of this platform as an efficient source UV-Vis frequency combs.

\begin{figure*}[t]
\centering\includegraphics[width=\linewidth]{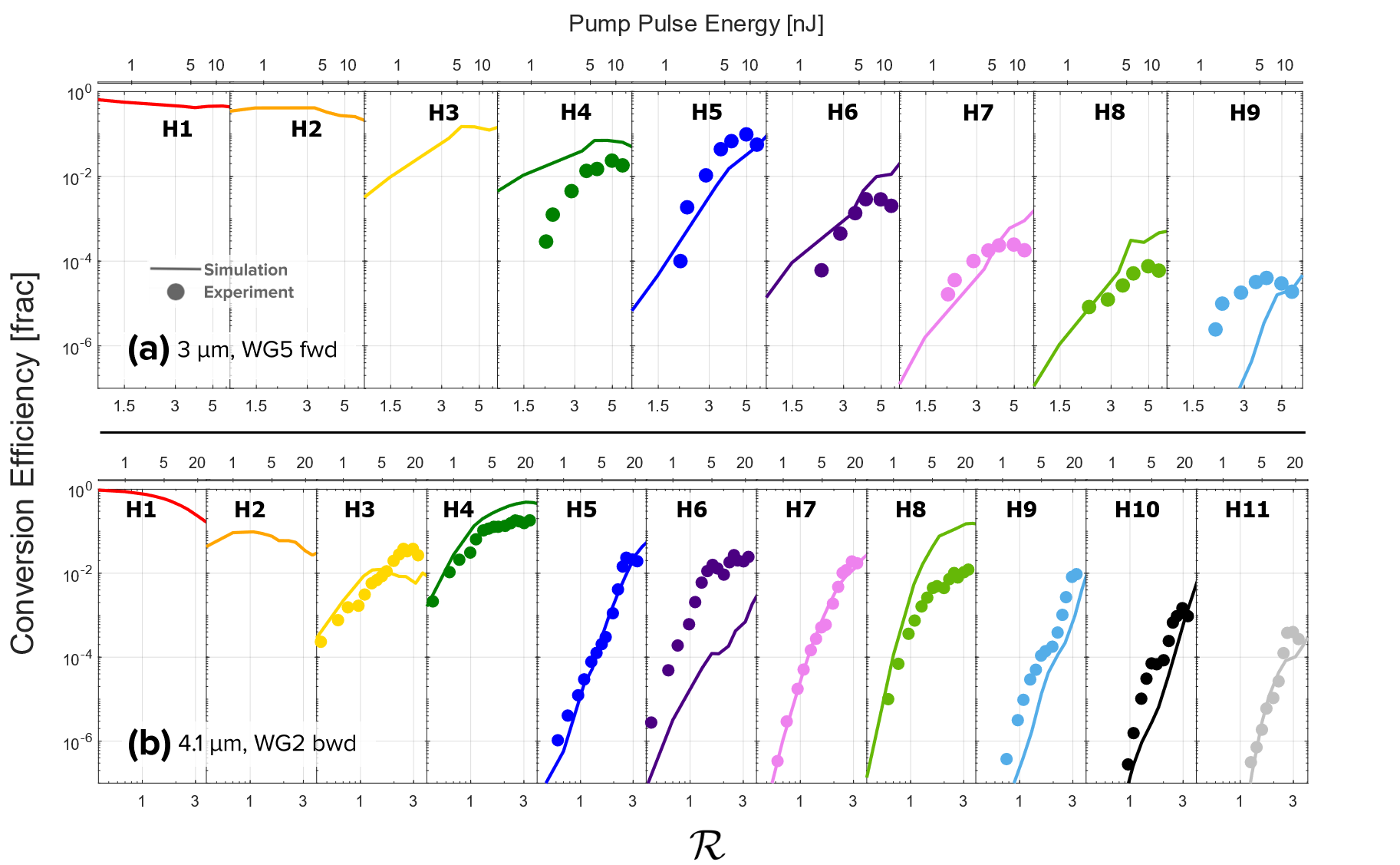}
\caption{Harmonic power scaling in terms of both the simulated $\mathcal{R} \equiv \Lgvm/\Lnl$ (bottom axes, lines) and experimental pump-pulse energy (top axes, dots) for \textbf{(a)} 3$\upmu$m, waveguide 5, forward, and  for \textbf{(b)} 4.1$\upmu$m, waveguide 2, backward, with 4.1 $\upmu$m experimental data from \cite{Hickstein_Optica2017}. Both experiments show saturation of harmonics at $\mathcal{R} \approx 3$.} 

\label{fig:powscale}
\end{figure*}


\subsection{Theoretical model}


We model the HHG processes using a 1$+$1D approach based on the single-mode nonlinear analytical envelope equation (NAEE) \cite{Conforti_PRA2010, Conforti_IEEE2010}, accounting for quadratic nonlinearity, dispersion, absorption, and self-steepening. The NAEE so far has been previously applied primarily to supercontinuum generation \cite{Phillips_OptExp2011}, conventional harmonic generation \cite{Conforti_PRA2010}, and broadband difference frequency generation \cite{Kowligy_OptLett2018, Lind_PRL2020}, but to our knowledge this is the first publication applying it to HHG. To emphasize the important length scales for cascaded HHG, we first recast the NAEE in the dimensionless form:

\begin{multline}
\frac{\partial U}{\partial z'} + iL_{\text{GVM}}\hat{D}U \\
= -i \ \text{sgn}(z) \frac{1}{2} \frac{L_{\text{GVM}}}{\Lnl}  \Big [ 1 - \frac{i}{\omega_0} \frac{\partial}{\partial \tau}  \Big ][U^2e^{i \varphi} + 2|U|^2e^{-i \varphi}] 
\label{ConfortiEq}
\end{multline}

where $U$ is a dimensionless field amplitude \cite{Agrawal_NonlinearFiberOpticsBook}, $\omega_0$ is the fundamental driver frequency, $\varphi = i \omega_0 \tau - i(\beta_0 - \beta_1 \omega_0)z$, $\tau = t - \beta_1z$ is the retarded time in a reference frame that moves with the group velocity of the fundamental, and $\hat{D}$ is the dispersion operator $\hat{D} = \sum\limits_{m=2}^{\infty} \frac{1}{m!} \beta_m \Big (-i \frac{\partial}{\partial \tau} \Big )^m$. Periodic poling is included explicitly via the factor $\text{sgn}(z) \equiv \text{sgn}(\cos{\frac{2\pi}{\Lambda}z})$, where $\Lambda(z)$ is the local poling period and $z$ is the forward propagation coordinate. The relative magnitude of dispersion and nonlinearity are expressed in terms of the length scales $\Lgvm$ and $\Lnl$ which we define as

\begin{equation}
\Lgvm \equiv |\beta_{1,H1} - \beta_{1,H2}|^{-1}\tau_{0}
\end{equation}

\begin{equation}
\Lnl \equiv \frac{n_0 c}{\omega_0 d_{\text{33}}|A_0|}
\end{equation}

where $n_0$ is the material index of refraction at the pump frequency, $d_{33}$ is the nonlinear coefficient, and $|A_0|$ is the peak field amplitude of the driver pulse in the medium. The pulse duration $\tau_0$ is defined as the full width at half maximum of the pulse divided by 1.76 \cite{Agrawal_NonlinearFiberOpticsBook}, and $\beta_{1,H1}$ and $\beta_{1,H2}$ are the inverse of the group velocities at the fundamental and second harmonic, respectively. Physically, $\Lgvm$ is the length scale over which two of the adjacent lower-order harmonics, here represented by the fundamental and second harmonic, walk off due to material dispersion, and $\Lnl$ is the characteristic length scale over which second harmonic generation would be expected to saturate with perfect phase matching in the absence of other processes \cite{Boyd:2003}. Length in equation (\ref{ConfortiEq}) is normalized via $z' \equiv z/L_{\text{GVM}}$. Increasing the pump power in the experiment corresponds to increasing the dimensionless ratio $\mathcal{R} \equiv \Lgvm/\Lnl$,
and thus the relative importance of the nonlinear source term in equation (\ref{ConfortiEq}). 

\ind For determining $\beta(\omega)$, and the resulting $\beta_n$ expansion coefficients, we use a calculation including waveguide dispersion from the COMSOL software package \cite{COMSOL_Ref}, although waveguide dispersion makes only a minor contribution. From this $\beta(\omega)$, we calculate $\Lgvm = $ 0.96 mm for the 3 $\upmu$m pump, whereas for the 4.1 $\upmu$m pump $\Lgvm =$ 0.49 mm due to the dispersion profile of lithium niobate. Over the range of our (measured) pump powers, using $d_{33} = 19.6$ pm/V \cite{Kowligy_OptLett2018, Shoji_JOSAB1997} and an effective area for the fundamental waveguide mode of 108 $\upmu$m$^2$, we estimate using the experimental parameters that $\Lnl$ ranges from $\sim$ 0.25 mm down to 85 $\upmu$m, approaching the scale of the waveguide poling period. In the simulations, we include absorption at ultraviolet frequencies above 967 THz and infrared frequencies below 60 THz, via error functions in the frequency-domain representation of the dispersion operator. 

\ind We solve equation (\ref{ConfortiEq}) using the embedded Runge-Kutta algorithm described in \cite{Balac_CompPhysComm2013}. The step size is determined by controlling the fractional error in each harmonic to a level of $<$ 0.1. This high accuracy tolerance results in typical step sizes less than 0.1 $\upmu$m. Furthermore, a large time grid and a high sampling rate are required to accurately represent the harmonics which spread over ranges of $\sim 100$ ps and $\sim$ 1000 THz simultaneously. This necessitates a time/frequency grid size of $2^{18}$ points, with a resulting computational cost for a single waveguide propagation of approximately 160 CPU core-hours. 

\ind Example output spectra for both 3 $\upmu$m and 4.1 $\upmu$m pumping are shown in figure \ref{fig:spectraoverlay}. The model reproduces the relatively smooth measured spectra and achieves reasonable agreement with both experiments for the relative magnitude of the harmonics. Similar agreement is observed for different waveguides and pump powers. In figure \ref{fig:powscale}, we compare the conversion efficiency of the model with experimental data for both 3 $\upmu$m pumping waveguide 5 in the forward direction (Fig. \ref{fig:powscale}a) and 4.1 $\upmu$m pumping waveguide 2 in the backward direction (Fig. \ref{fig:powscale}b). In matching theory with experiment the only free parameter used is the relative scaling between the experimental pump power axis and the values of $\mathcal{R}$ used in the simulations, such that the simulations are effectively performed with a pump pulse energy 3.8 (1.6) times lower those estimated for the 3 (4.1) $\upmu$m experiments. We note that such adjustments are common when comparing simulation and experiment in waveguide nonlinear optics \cite{Couch_Optica2020, Kowligy_OptLett2018, Phillips_OptLett2011_2}. Viewed another way, to the extent that the 1+1D simulation captures the physics of the HHG process, overlaying the simulation results with experiments this way finds the parameters (i.e. power coupled into the waveguide fundamental mode) with which the experiments are actually done. Once this adjustment is made, experiment and theory agree within approximately 1 order of magnitude over a wide range of parameters.

\ind Before saturation, naively one might expect the $q^{th}$-harmonic power for a process based on perturbative nonlinear optics to scale as the the pump power to the $q^{th}$ power. However, as in Hickstein et al. \cite{Hickstein_Optica2017}, even in the threshold regime (i.e. before saturation) we find the harmonic power grows much more slowly than $(E_{\textss{pulse}})^q$, or conversion efficiency that grows as $(E_{\textss{pulse}})^{(q-1)}$. We find instead that power-law fits to the data in figure \ref{fig:powscale} give exponents in the range of 0.5 to 6 and always less than $(E_{\textss{pulse}})^{(q-1)}$. These slopes on the log-log plot are reproduced reasonably well by the simulations, particularly for the 4.1 $\upmu$m experiment which recorded more data in the threshold regime. 

\ind We find that the high-order harmonics in the visible spectral range begin to reach saturation for both 3 and 4.1 $\upmu$m pumping around $\mathcal{R} \approx 3$. This has an intuitive interpretation. Since the mixing of many harmonics is required to achieve cascaded HHG, three-wave mixing must be saturated several times over before group-velocity mismatch separates the participating harmonics. The near-universal shape of the saturated HHG spectra of figure \ref{fig:spectra}a and robustness of this $\mathcal{R} \approx 3$ result across different waveguides, poling period chirp directions, and pump pulse center wavelengths, suggests it to be a generally applicable metric for evaluating cascaded HHG physics.

\begin{figure}[t]
\centering\includegraphics[angle = 90, origin = c, width=0.97\linewidth]{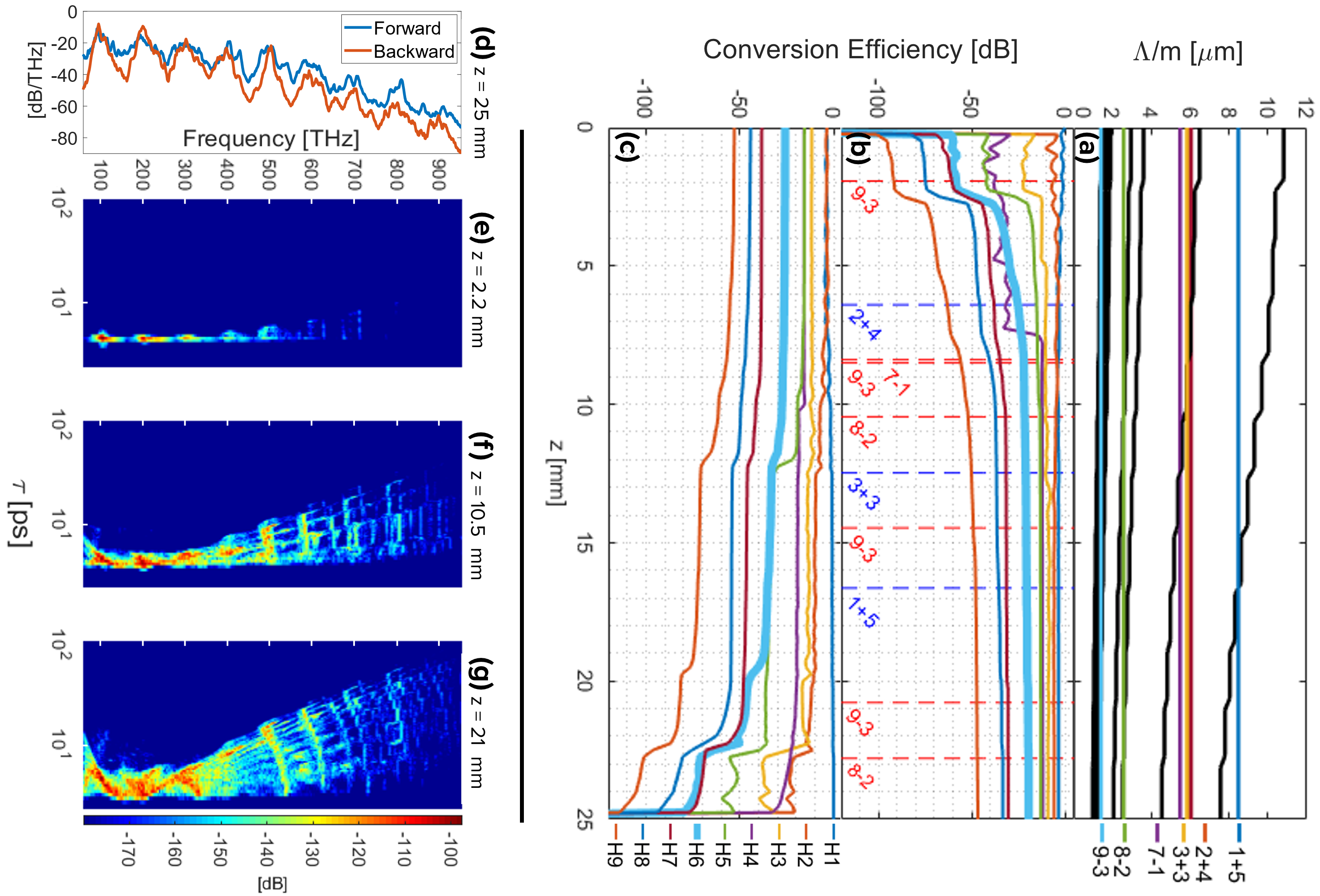}
\caption{
\textbf{(a)} Higher-order poling periods of Waveguide 5 (black) intersected by the phase-matching solutions (horizontal colored) of SFG/DFG processes generating the 6th-harmonic of 3 $\upmu$m. \textbf{(b, c)} Harmonic evolution throughout propagation in the forward and backward pumping direction for the same $\mathcal{R}$ as figure \ref{fig:spectraoverlay}a, overlaid with the phase-matched locations of SFG (blue dashed) and DFG (red dashed) processes generating H6 (thick blue line). While the QPM direction indeed alters each harmonic's growth history, ultimately their conversion efficiencies are similar, further highlighted in \textbf{(d)} by the forward (blue) and backward (orange) output spectra. \textbf{(e-g)} Time-frequency spectrograms corresponding to (b) at three waveguide locations, showing the dispersion of generated harmonics and their temporal-overlap throughout propagation.
}
\label{fig:pathways}
\end{figure}

%

\subsection{Harmonic pathways and quasi-phase-matching}

Unlike the experiment which measures only the output of a waveguide, the simulation provides information about the optical fields at all points along the waveguide (Fig. \ref{fig:pathways}). By monitoring each harmonic throughout propagation in either waveguide direction, we can uncover QPM dependencies and correlate a harmonic's evolving position and growth with that of other harmonics, and therefore attempt to assign generation pathways based on specific sum-frequency generation (SFG) or difference-frequency generation (DFG) processes. The generation of any harmonic by a given SFG or DFG process is essentially a threefold-constrained problem: 1) the process must be phase matched at some poling period of the waveguide (Fig. \ref{fig:pathways}a), 2) the participating fields must already exist with sufficient power for efficient nonlinear interaction (i.e. after generation or before consumption by some other process) (Fig. \ref{fig:pathways}b,c), and 3) the interacting fields must be temporally overlapped at the locations where 1) and 2) are satisfied prior to walk-off  (Fig. \ref{fig:pathways}e-g). 

\ind Such a collision of circumstances in this 3D parameter space would at first seem unlikely, but the large number of simultaneous constraints is compensated by the large number of possibilities to satisfy them. First there are higher orders of the quasi-phase matching conditions, such that a three-wave mixing process is not only satisfied at $\Delta k$ = $2\pi/\Lambda$, but also any multiple $m (2\pi/\Lambda)$, where $m$ is an odd integer. Viewed in real space, a process can be quasi-phase matched using sub-harmonics of the poling period $\Lambda/m$, as illustrated in figure \ref{fig:pathways}a. Second is the fact that there are many possible three-wave mixing pathways to generate a particular harmonic. This large number of possibilities makes a global analysis of the impact of the periodic poling profiles difficult.

\ind We present here a case study of harmonic 6 in waveguide 5, pumped by 3 $\upmu$m light in both the forward and backward directions. The SFG and DFG processes that can generate harmonic 6 are
\begin{align*}
H6 = H1 + H5  &&  H6 = H9 - H3 \\
H6 = H2 + H4  &&  H6 = H8 - H2 \\
H6 = H3 + H3  &&  H6 = H7 - H1 
\end{align*}
Positions where these processes are phase matched occur throughout the waveguide and are shown in figure \ref{fig:pathways}b. In general, we find both SFG and DFG processes to be important in cascaded HHG, as also concluded by Park et al. \cite{Park_OptLett2017} for the case of fractal poling. However, first SFG must build those higher-order harmonics which later combine via DFG to generate lower-order harmonics. For this reason the direction of the waveguide matters, affecting whether first primarily the lower-harmonic (longer poling periods) or high-harmonic (shorter poling period) processes are phase matched. Depicted here in figure \ref{fig:pathways}b,c, pumping waveguide 5 in either the forward or backward direction initiates a different growth history for each harmonic. For instance, whereas in the forward direction H2 is immediately generated and offers an intense partner for SFG and DFG processes throughout the length of the waveguide, in the backward direction it builds up more slowly. Ultimately, however, in either pumping direction, by the end of the waveguide the conversion efficiency to any harmonic is similar (Fig. \ref{fig:pathways}d). At least one of the aforementioned 3D collisions happens in both directions.


\ind In this presented case study and in most cases, we find that most high harmonic generation occurs (Fig. \ref{fig:pathways}b,c) in waveguide regions where a large number of processes can be phase matched within a distance shorter than $\Lgvm$. This can be further aided by chirping the poling period more rapidly. Also, temporal compression of the pump pulse within the waveguide can be significant, suggesting that the input pulse chirp may be tuned to position this enhancement at certain poling locations.
 
\subsection{Discussion}

While the agreement level between the 1+1D theory and experiment presented here is sufficient to identify cascaded $\chi^{(2)}$ processes as the dominant mechanism for HHG in the waveguides, significant discrepancies remain. We briefly discuss avenues to improve the model. 

\ind Considering the modest $\sim$ 100 kW peak pump powers, we may dismiss self-focusing effects. Furthermore, although significant self-phase modulation is observed on the fundamental (Fig. \ref{fig:setup}a), we found inclusion of a Kerr-effect term in equation (\ref{ConfortiEq}), in similar fashion to \cite{Kowligy_OptLett2018}, to produce negligible effects on the HHG.

\ind For the 3 $\upmu$m wavelength, the waveguide supports 3 modes, with the number of supported modes for the harmonics scaling as $q^2$. A more sophisticated model would include multi-mode effects \cite{Phillips_OptExp2011, Liu_OptLett2016}. The launched fundamental may couple to multiple modes, and also harmonics generated by the fundamental can populate multiple modes. The impact of including higher-order modes in the model is unclear. On the one hand, we estimate the extra phase mismatch due to waveguide dispersion in the higher-order modes to be negligible, meaning that light in higher-order modes can readily mix with fundamental modes and produce interference. However, we also estimate the nonlinear coupling between modes to be weak. For example, we calculated the overlap integral of the the product $E_{1,00} E_{2,00}$ with $E_{3,00}$ and find it to be greater than 0.96, suggesting that the single-mode envelope equation (\ref{ConfortiEq}) should capture most of the wave mixing physics. While the harmonics appeared (visibly) to emerge in a TEM$_{00}$-like beam, the amount of fundamental light launched in higher-order waveguide modes is not known and this could have a significant impact on modeling. Also interference between light in different waveguide modes could help explain the sharp features of the harmonic spectra.

\ind Another factor to consider in improving the models is the dispersion of the nonlinear susceptibility \cite{Shoji_JOSAB1997, dorozhkin_dispersion_1976}, which is neglected here where we treat $\chi^{(2)}$ as a constant, independent of frequency. Over the range of our harmonic spectra, dispersion of $\chi^{(2)}$ may be significant, particularly for the mixing of higher-order ultraviolet harmonics where  $\chi^{(2)}$ is enhanced by as much as a factor of 3 \cite{Shoji_JOSAB1997, dorozhkin_dispersion_1976}. DFG processes involving the higher orders are relevant for the HHG process, and we suspect including the frequency dependence of $\chi^{(2)}$, via a time-dependent nonlinear response function \cite{Agrawal_NonlinearFiberOpticsBook}, could make a significant improvement in the results.

\section{Conclusion}

In this article, we report efficient high-order harmonic generation from chirped periodically-poled lithium niobate waveguides driven with 3 $\upmu$m pulses with up to 12 nJ of pump pulse energy at 100 MHz. High-brightness HHG frequency combs are observed emerging from the waveguide covering the visible and UV up to the lithium niobate band gap, with generated powers greater than 100 mW/harmonic. Importantly, we verified that the harmonic emissions in cascaded HHG inherit the coherence properties of the driving frequency comb despite the highly nonlinear process. 

\ind Semi-quantitative agreement with simulations identifies cascaded HHG as the dominant generation mechanism. Furthermore, comparison with simulations points to a critical dimensionless parameter $\mathcal{R} \equiv \Lgvm/\Lnl$ that determines the turn-on and saturation of the cascaded HHG process. With these understandings, future work may explore more sophisticated schemes for enhanced conversion efficiencies and tailored harmonic spectra via optimization of the waveguide poling period and pump pulse parameters.

\smallskip

\textbf{Funding} National Science Foundation (NSF) (1708743); Air
Force Office of Scientific Research (AFOSR) (FA9550-16-1-0016, FA9550-16-1-0164, and FA9550-20-1-0259). Defense Advanced Research Projects Agency (DARPA) SCOUT program. National Institute of Standards and Technology NIST on a Chip. A. Catanese acknowledges support from the GAANN program of the U.S. Dept. of Education. Mention of specific products or trade names does not constitute an endorsement by NIST. 

\smallskip

\textbf{Disclosures} The authors declare no conflicts of interest.


\begin{thebibliography}{41}%
\makeatletter
\providecommand \@ifxundefined [1]{%
 \@ifx{#1\undefined}
}%
\providecommand \@ifnum [1]{%
 \ifnum #1\expandafter \@firstoftwo
 \else \expandafter \@secondoftwo
 \fi
}%
\providecommand \@ifx [1]{%
 \ifx #1\expandafter \@firstoftwo
 \else \expandafter \@secondoftwo
 \fi
}%
\providecommand \natexlab [1]{#1}%
\providecommand \enquote  [1]{``#1''}%
\providecommand \bibnamefont  [1]{#1}%
\providecommand \bibfnamefont [1]{#1}%
\providecommand \citenamefont [1]{#1}%
\providecommand \href@noop [0]{\@secondoftwo}%
\providecommand \href [0]{\begingroup \@sanitize@url \@href}%
\providecommand \@href[1]{\@@startlink{#1}\@@href}%
\providecommand \@@href[1]{\endgroup#1\@@endlink}%
\providecommand \@sanitize@url [0]{\catcode `\\12\catcode `\$12\catcode
  `\&12\catcode `\#12\catcode `\^12\catcode `\_12\catcode `\%12\relax}%
\providecommand \@@startlink[1]{}%
\providecommand \@@endlink[0]{}%
\providecommand \url  [0]{\begingroup\@sanitize@url \@url }%
\providecommand \@url [1]{\endgroup\@href {#1}{\urlprefix }}%
\providecommand \urlprefix  [0]{URL }%
\providecommand \Eprint [0]{\href }%
\providecommand \doibase [0]{https://doi.org/}%
\providecommand \selectlanguage [0]{\@gobble}%
\providecommand \bibinfo  [0]{\@secondoftwo}%
\providecommand \bibfield  [0]{\@secondoftwo}%
\providecommand \translation [1]{[#1]}%
\providecommand \BibitemOpen [0]{}%
\providecommand \bibitemStop [0]{}%
\providecommand \bibitemNoStop [0]{.\EOS\space}%
\providecommand \EOS [0]{\spacefactor3000\relax}%
\providecommand \BibitemShut  [1]{\csname bibitem#1\endcsname}%
\let\auto@bib@innerbib\@empty
\bibitem [{\citenamefont {Ferray}\ \emph {et~al.}(1988)\citenamefont {Ferray},
  \citenamefont {L'Huillier}, \citenamefont {Li}, \citenamefont {Lompre},
  \citenamefont {Mainfray},\ and\ \citenamefont {Manus}}]{Ferray_JPhysB1988}%
  \BibitemOpen
  \bibfield  {author} {\bibinfo {author} {\bibfnamefont {M.}~\bibnamefont
  {Ferray}}, \bibinfo {author} {\bibfnamefont {A.}~\bibnamefont {L'Huillier}},
  \bibinfo {author} {\bibfnamefont {X.~F.}\ \bibnamefont {Li}}, \bibinfo
  {author} {\bibfnamefont {L.~A.}\ \bibnamefont {Lompre}}, \bibinfo {author}
  {\bibfnamefont {G.}~\bibnamefont {Mainfray}},\ and\ \bibinfo {author}
  {\bibfnamefont {C.}~\bibnamefont {Manus}},\ }\bibfield  {title} {\bibinfo
  {title} {Multiple-harmonic conversion of 1064 nm radiation in rare gases},\
  }\href {http://stacks.iop.org/0953-4075/21/i=3/a=001} {\bibfield  {journal}
  {\bibinfo  {journal} {Journal of Physics B: Atomic, Molecular and Optical
  Physics}\ }\textbf {\bibinfo {volume} {21}},\ \bibinfo {pages} {L31}
  (\bibinfo {year} {1988})}\BibitemShut {NoStop}%
\bibitem [{\citenamefont {McPherson}\ \emph {et~al.}(1987)\citenamefont
  {McPherson}, \citenamefont {Gibson}, \citenamefont {Jara}, \citenamefont
  {Johann}, \citenamefont {Luk}, \citenamefont {McIntyre}, \citenamefont
  {Boyer},\ and\ \citenamefont {Rhodes}}]{McPherson_JOSAB1987}%
  \BibitemOpen
  \bibfield  {author} {\bibinfo {author} {\bibfnamefont {A.}~\bibnamefont
  {McPherson}}, \bibinfo {author} {\bibfnamefont {G.}~\bibnamefont {Gibson}},
  \bibinfo {author} {\bibfnamefont {H.}~\bibnamefont {Jara}}, \bibinfo {author}
  {\bibfnamefont {U.}~\bibnamefont {Johann}}, \bibinfo {author} {\bibfnamefont
  {T.~S.}\ \bibnamefont {Luk}}, \bibinfo {author} {\bibfnamefont {I.~A.}\
  \bibnamefont {McIntyre}}, \bibinfo {author} {\bibfnamefont {K.}~\bibnamefont
  {Boyer}},\ and\ \bibinfo {author} {\bibfnamefont {C.~K.}\ \bibnamefont
  {Rhodes}},\ }\bibfield  {title} {\bibinfo {title} {Studies of multiphoton
  production of vacuum-ultraviolet radiation in the rare gases},\ }\href
  {https://doi.org/10.1364/JOSAB.4.000595} {\bibfield  {journal} {\bibinfo
  {journal} {J. Opt. Soc. Am. B}\ }\textbf {\bibinfo {volume} {4}},\ \bibinfo
  {pages} {595} (\bibinfo {year} {1987})}\BibitemShut {NoStop}%
\bibitem [{\citenamefont {Ghimire}\ \emph {et~al.}(2011)\citenamefont
  {Ghimire}, \citenamefont {DiChiara}, \citenamefont {Sistrunk}, \citenamefont
  {Agostini}, \citenamefont {DiMauro},\ and\ \citenamefont
  {Reis}}]{Ghimire_NatPhys2011}%
  \BibitemOpen
  \bibfield  {author} {\bibinfo {author} {\bibfnamefont {S.}~\bibnamefont
  {Ghimire}}, \bibinfo {author} {\bibfnamefont {A.~D.}\ \bibnamefont
  {DiChiara}}, \bibinfo {author} {\bibfnamefont {E.}~\bibnamefont {Sistrunk}},
  \bibinfo {author} {\bibfnamefont {P.}~\bibnamefont {Agostini}}, \bibinfo
  {author} {\bibfnamefont {L.~F.}\ \bibnamefont {DiMauro}},\ and\ \bibinfo
  {author} {\bibfnamefont {D.~A.}\ \bibnamefont {Reis}},\ }\bibfield  {title}
  {\bibinfo {title} {Observation of high-order harmonic generation in a bulk
  crystal},\ }\href {http://dx.doi.org/10.1038/nphys1847} {\bibfield  {journal}
  {\bibinfo  {journal} {Nat Phys}\ }\textbf {\bibinfo {volume} {7}},\ \bibinfo
  {pages} {138} (\bibinfo {year} {2011})}\BibitemShut {NoStop}%
\bibitem [{\citenamefont {DiChiara}\ \emph {et~al.}(2009)\citenamefont
  {DiChiara}, \citenamefont {Sistrunk}, \citenamefont {Miller}, \citenamefont
  {Agostini},\ and\ \citenamefont {DiMauro}}]{DiChiara_OptExp2009}%
  \BibitemOpen
  \bibfield  {author} {\bibinfo {author} {\bibfnamefont {A.~D.}\ \bibnamefont
  {DiChiara}}, \bibinfo {author} {\bibfnamefont {E.}~\bibnamefont {Sistrunk}},
  \bibinfo {author} {\bibfnamefont {T.~A.}\ \bibnamefont {Miller}}, \bibinfo
  {author} {\bibfnamefont {P.}~\bibnamefont {Agostini}},\ and\ \bibinfo
  {author} {\bibfnamefont {L.~F.}\ \bibnamefont {DiMauro}},\ }\bibfield
  {title} {\bibinfo {title} {An investigation of harmonic generation in liquid
  media with a mid-infrared laser},\ }\href
  {https://doi.org/10.1364/OE.17.020959} {\bibfield  {journal} {\bibinfo
  {journal} {Opt. Express}\ }\textbf {\bibinfo {volume} {17}},\ \bibinfo
  {pages} {20959} (\bibinfo {year} {2009})}\BibitemShut {NoStop}%
\bibitem [{\citenamefont {Luu}\ \emph {et~al.}(2018)\citenamefont {Luu},
  \citenamefont {Yin}, \citenamefont {Jain}, \citenamefont {Gaumnitz},
  \citenamefont {Pertot}, \citenamefont {Ma},\ and\ \citenamefont
  {W{\"o}rner}}]{Luu_NatComm2018}%
  \BibitemOpen
  \bibfield  {author} {\bibinfo {author} {\bibfnamefont {T.~T.}\ \bibnamefont
  {Luu}}, \bibinfo {author} {\bibfnamefont {Z.}~\bibnamefont {Yin}}, \bibinfo
  {author} {\bibfnamefont {A.}~\bibnamefont {Jain}}, \bibinfo {author}
  {\bibfnamefont {T.}~\bibnamefont {Gaumnitz}}, \bibinfo {author}
  {\bibfnamefont {Y.}~\bibnamefont {Pertot}}, \bibinfo {author} {\bibfnamefont
  {J.}~\bibnamefont {Ma}},\ and\ \bibinfo {author} {\bibfnamefont {H.~J.}\
  \bibnamefont {W{\"o}rner}},\ }\bibfield  {title} {\bibinfo {title}
  {Extreme--ultraviolet high--harmonic generation in liquids},\ }\href
  {https://doi.org/10.1038/s41467-018-06040-4} {\bibfield  {journal} {\bibinfo
  {journal} {Nature Communications}\ }\textbf {\bibinfo {volume} {9}},\
  \bibinfo {pages} {3723} (\bibinfo {year} {2018})}\BibitemShut {NoStop}%
\bibitem [{\citenamefont {Dromey}\ \emph {et~al.}(2006)\citenamefont {Dromey},
  \citenamefont {Zepf}, \citenamefont {Gopal}, \citenamefont {Lancaster},
  \citenamefont {Wei}, \citenamefont {Krushelnick}, \citenamefont {Tatarakis},
  \citenamefont {Vakakis}, \citenamefont {Moustaizis}, \citenamefont {Kodama},
  \citenamefont {Tampo}, \citenamefont {Stoeckl}, \citenamefont {Clarke},
  \citenamefont {Habara}, \citenamefont {Neely}, \citenamefont {Karsch},\ and\
  \citenamefont {Norreys}}]{Dromey_NatPhys2006}%
  \BibitemOpen
  \bibfield  {author} {\bibinfo {author} {\bibfnamefont {B.}~\bibnamefont
  {Dromey}}, \bibinfo {author} {\bibfnamefont {M.}~\bibnamefont {Zepf}},
  \bibinfo {author} {\bibfnamefont {A.}~\bibnamefont {Gopal}}, \bibinfo
  {author} {\bibfnamefont {K.}~\bibnamefont {Lancaster}}, \bibinfo {author}
  {\bibfnamefont {M.~S.}\ \bibnamefont {Wei}}, \bibinfo {author} {\bibfnamefont
  {K.}~\bibnamefont {Krushelnick}}, \bibinfo {author} {\bibfnamefont
  {M.}~\bibnamefont {Tatarakis}}, \bibinfo {author} {\bibfnamefont
  {N.}~\bibnamefont {Vakakis}}, \bibinfo {author} {\bibfnamefont
  {S.}~\bibnamefont {Moustaizis}}, \bibinfo {author} {\bibfnamefont
  {R.}~\bibnamefont {Kodama}}, \bibinfo {author} {\bibfnamefont
  {M.}~\bibnamefont {Tampo}}, \bibinfo {author} {\bibfnamefont
  {C.}~\bibnamefont {Stoeckl}}, \bibinfo {author} {\bibfnamefont
  {R.}~\bibnamefont {Clarke}}, \bibinfo {author} {\bibfnamefont
  {H.}~\bibnamefont {Habara}}, \bibinfo {author} {\bibfnamefont
  {D.}~\bibnamefont {Neely}}, \bibinfo {author} {\bibfnamefont
  {S.}~\bibnamefont {Karsch}},\ and\ \bibinfo {author} {\bibfnamefont
  {P.}~\bibnamefont {Norreys}},\ }\bibfield  {title} {\bibinfo {title} {High
  harmonic generation in the relativistic limit},\ }\href
  {http://dx.doi.org/10.1038/nphys338} {\bibfield  {journal} {\bibinfo
  {journal} {Nat Phys}\ }\textbf {\bibinfo {volume} {2}},\ \bibinfo {pages}
  {456} (\bibinfo {year} {2006})}\BibitemShut {NoStop}%
\bibitem [{\citenamefont {Kandula}\ \emph {et~al.}(2010)\citenamefont
  {Kandula}, \citenamefont {Gohle}, \citenamefont {Pinkert}, \citenamefont
  {Ubachs},\ and\ \citenamefont {Eikema}}]{Kandula_PRL2010}%
  \BibitemOpen
  \bibfield  {author} {\bibinfo {author} {\bibfnamefont {D.~Z.}\ \bibnamefont
  {Kandula}}, \bibinfo {author} {\bibfnamefont {C.}~\bibnamefont {Gohle}},
  \bibinfo {author} {\bibfnamefont {T.~J.}\ \bibnamefont {Pinkert}}, \bibinfo
  {author} {\bibfnamefont {W.}~\bibnamefont {Ubachs}},\ and\ \bibinfo {author}
  {\bibfnamefont {K.~S.~E.}\ \bibnamefont {Eikema}},\ }\bibfield  {title}
  {\bibinfo {title} {Extreme ultraviolet frequency comb metrology},\ }\href
  {https://doi.org/10.1103/PhysRevLett.105.063001} {\bibfield  {journal}
  {\bibinfo  {journal} {Phys. Rev. Lett.}\ }\textbf {\bibinfo {volume} {105}},\
  \bibinfo {pages} {063001} (\bibinfo {year} {2010})}\BibitemShut {NoStop}%
\bibitem [{\citenamefont {Cing\"{o}z}\ \emph {et~al.}(2012)\citenamefont
  {Cing\"{o}z}, \citenamefont {Yost}, \citenamefont {Allison}, \citenamefont
  {Ruehl}, \citenamefont {Fermann}, \citenamefont {Hartl},\ and\ \citenamefont
  {Ye}}]{Cingoz_Nature2012}%
  \BibitemOpen
  \bibfield  {author} {\bibinfo {author} {\bibfnamefont {A.}~\bibnamefont
  {Cing\"{o}z}}, \bibinfo {author} {\bibfnamefont {D.~C.}\ \bibnamefont
  {Yost}}, \bibinfo {author} {\bibfnamefont {T.~K.}\ \bibnamefont {Allison}},
  \bibinfo {author} {\bibfnamefont {A.}~\bibnamefont {Ruehl}}, \bibinfo
  {author} {\bibfnamefont {M.~E.}\ \bibnamefont {Fermann}}, \bibinfo {author}
  {\bibfnamefont {I.}~\bibnamefont {Hartl}},\ and\ \bibinfo {author}
  {\bibfnamefont {J.}~\bibnamefont {Ye}},\ }\bibfield  {title} {\bibinfo
  {title} {Direct frequency comb spectroscopy in the extreme ultraviolet},\
  }\href {http://dx.doi.org/10.1038/nature10711} {\bibfield  {journal}
  {\bibinfo  {journal} {Nature}\ }\textbf {\bibinfo {volume} {482}},\ \bibinfo
  {pages} {68} (\bibinfo {year} {2012})}\BibitemShut {NoStop}%
\bibitem [{\citenamefont {Allison}\ \emph {et~al.}(2010)\citenamefont
  {Allison}, \citenamefont {Wright}, \citenamefont {Stooke}, \citenamefont
  {Khurmi}, \citenamefont {van Tilborg}, \citenamefont {Liu}, \citenamefont
  {Falcone},\ and\ \citenamefont {Belkacem}}]{Allison_OptLett2010}%
  \BibitemOpen
  \bibfield  {author} {\bibinfo {author} {\bibfnamefont {T.~K.}\ \bibnamefont
  {Allison}}, \bibinfo {author} {\bibfnamefont {T.~W.}\ \bibnamefont {Wright}},
  \bibinfo {author} {\bibfnamefont {A.~M.}\ \bibnamefont {Stooke}}, \bibinfo
  {author} {\bibfnamefont {C.}~\bibnamefont {Khurmi}}, \bibinfo {author}
  {\bibfnamefont {J.}~\bibnamefont {van Tilborg}}, \bibinfo {author}
  {\bibfnamefont {Y.}~\bibnamefont {Liu}}, \bibinfo {author} {\bibfnamefont
  {R.~W.}\ \bibnamefont {Falcone}},\ and\ \bibinfo {author} {\bibfnamefont
  {A.}~\bibnamefont {Belkacem}},\ }\bibfield  {title} {\bibinfo {title}
  {Femtosecond spectroscopy with vacuum ultraviolet pulse pairs},\ }\href
  {https://doi.org/10.1364/OL.35.003664} {\bibfield  {journal} {\bibinfo
  {journal} {Opt. Lett.}\ }\textbf {\bibinfo {volume} {35}},\ \bibinfo {pages}
  {3664} (\bibinfo {year} {2010})}\BibitemShut {NoStop}%
\bibitem [{\citenamefont {Corder}\ \emph {et~al.}(2018)\citenamefont {Corder},
  \citenamefont {Zhao}, \citenamefont {Bakalis}, \citenamefont {Li},
  \citenamefont {Kershis}, \citenamefont {Muraca}, \citenamefont {White},\ and\
  \citenamefont {Allison}}]{Corder_StructDyn2018}%
  \BibitemOpen
  \bibfield  {author} {\bibinfo {author} {\bibfnamefont {C.}~\bibnamefont
  {Corder}}, \bibinfo {author} {\bibfnamefont {P.}~\bibnamefont {Zhao}},
  \bibinfo {author} {\bibfnamefont {J.}~\bibnamefont {Bakalis}}, \bibinfo
  {author} {\bibfnamefont {X.}~\bibnamefont {Li}}, \bibinfo {author}
  {\bibfnamefont {M.~D.}\ \bibnamefont {Kershis}}, \bibinfo {author}
  {\bibfnamefont {A.~R.}\ \bibnamefont {Muraca}}, \bibinfo {author}
  {\bibfnamefont {M.~G.}\ \bibnamefont {White}},\ and\ \bibinfo {author}
  {\bibfnamefont {T.~K.}\ \bibnamefont {Allison}},\ }\bibfield  {title}
  {\bibinfo {title} {Ultrafast extreme ultraviolet photoemission without space
  charge},\ }\href {https://doi.org/10.1063/1.5045578} {\bibfield  {journal}
  {\bibinfo  {journal} {Structural Dynamics}\ }\textbf {\bibinfo {volume}
  {5}},\ \bibinfo {pages} {054301} (\bibinfo {year} {2018})},\ \Eprint
  {https://arxiv.org/abs/https://doi.org/10.1063/1.5045578}
  {https://doi.org/10.1063/1.5045578} \BibitemShut {NoStop}%
\bibitem [{\citenamefont {Sandberg}\ \emph {et~al.}(2007)\citenamefont
  {Sandberg}, \citenamefont {Paul}, \citenamefont {Raymondson}, \citenamefont
  {H\"adrich}, \citenamefont {Gaudiosi}, \citenamefont {Holtsnider},
  \citenamefont {Tobey}, \citenamefont {Cohen}, \citenamefont {Murnane},
  \citenamefont {Kapteyn}, \citenamefont {Song}, \citenamefont {Miao},
  \citenamefont {Liu},\ and\ \citenamefont {Salmassi}}]{Sandberg_PRL2007}%
  \BibitemOpen
  \bibfield  {author} {\bibinfo {author} {\bibfnamefont {R.~L.}\ \bibnamefont
  {Sandberg}}, \bibinfo {author} {\bibfnamefont {A.}~\bibnamefont {Paul}},
  \bibinfo {author} {\bibfnamefont {D.~A.}\ \bibnamefont {Raymondson}},
  \bibinfo {author} {\bibfnamefont {S.}~\bibnamefont {H\"adrich}}, \bibinfo
  {author} {\bibfnamefont {D.~M.}\ \bibnamefont {Gaudiosi}}, \bibinfo {author}
  {\bibfnamefont {J.}~\bibnamefont {Holtsnider}}, \bibinfo {author}
  {\bibfnamefont {R.~I.}\ \bibnamefont {Tobey}}, \bibinfo {author}
  {\bibfnamefont {O.}~\bibnamefont {Cohen}}, \bibinfo {author} {\bibfnamefont
  {M.~M.}\ \bibnamefont {Murnane}}, \bibinfo {author} {\bibfnamefont {H.~C.}\
  \bibnamefont {Kapteyn}}, \bibinfo {author} {\bibfnamefont {C.}~\bibnamefont
  {Song}}, \bibinfo {author} {\bibfnamefont {J.}~\bibnamefont {Miao}}, \bibinfo
  {author} {\bibfnamefont {Y.}~\bibnamefont {Liu}},\ and\ \bibinfo {author}
  {\bibfnamefont {F.}~\bibnamefont {Salmassi}},\ }\bibfield  {title} {\bibinfo
  {title} {Lensless diffractive imaging using tabletop coherent high-harmonic
  soft-x-ray beams},\ }\href {https://doi.org/10.1103/PhysRevLett.99.098103}
  {\bibfield  {journal} {\bibinfo  {journal} {Phys. Rev. Lett.}\ }\textbf
  {\bibinfo {volume} {99}},\ \bibinfo {pages} {098103} (\bibinfo {year}
  {2007})}\BibitemShut {NoStop}%
\bibitem [{\citenamefont {Mikkelsen}\ \emph {et~al.}(2009)\citenamefont
  {Mikkelsen}, \citenamefont {Schwenke}, \citenamefont {Fordell}, \citenamefont
  {Luo}, \citenamefont {Kl{\"u}nder}, \citenamefont {Hilner}, \citenamefont
  {Anttu}, \citenamefont {Zakharov}, \citenamefont {Lundgren}, \citenamefont
  {Mauritsson}, \citenamefont {Andersen}, \citenamefont {Xu},\ and\
  \citenamefont {L'Huillier}}]{Mikkelsen_RSI2009}%
  \BibitemOpen
  \bibfield  {author} {\bibinfo {author} {\bibfnamefont {A.}~\bibnamefont
  {Mikkelsen}}, \bibinfo {author} {\bibfnamefont {J.}~\bibnamefont {Schwenke}},
  \bibinfo {author} {\bibfnamefont {T.}~\bibnamefont {Fordell}}, \bibinfo
  {author} {\bibfnamefont {G.}~\bibnamefont {Luo}}, \bibinfo {author}
  {\bibfnamefont {K.}~\bibnamefont {Kl{\"u}nder}}, \bibinfo {author}
  {\bibfnamefont {E.}~\bibnamefont {Hilner}}, \bibinfo {author} {\bibfnamefont
  {N.}~\bibnamefont {Anttu}}, \bibinfo {author} {\bibfnamefont {A.~A.}\
  \bibnamefont {Zakharov}}, \bibinfo {author} {\bibfnamefont {E.}~\bibnamefont
  {Lundgren}}, \bibinfo {author} {\bibfnamefont {J.}~\bibnamefont
  {Mauritsson}}, \bibinfo {author} {\bibfnamefont {J.~N.}\ \bibnamefont
  {Andersen}}, \bibinfo {author} {\bibfnamefont {H.~Q.}\ \bibnamefont {Xu}},\
  and\ \bibinfo {author} {\bibfnamefont {A.}~\bibnamefont {L'Huillier}},\
  }\bibfield  {title} {\bibinfo {title} {Photoemission electron microscopy
  using extreme ultraviolet attosecond pulse trains},\ }\href
  {https://doi.org/http://dx.doi.org/10.1063/1.3263759} {\bibfield  {journal}
  {\bibinfo  {journal} {Review of Scientific Instruments}\ }\textbf {\bibinfo
  {volume} {80}},\ \bibinfo {eid} {123703} (\bibinfo {year}
  {2009})}\BibitemShut {NoStop}%
\bibitem [{\citenamefont {Chini}\ \emph {et~al.}(2014)\citenamefont {Chini},
  \citenamefont {Zhao},\ and\ \citenamefont {Chang}}]{Chini_NatPhot2014_2}%
  \BibitemOpen
  \bibfield  {author} {\bibinfo {author} {\bibfnamefont {M.}~\bibnamefont
  {Chini}}, \bibinfo {author} {\bibfnamefont {K.}~\bibnamefont {Zhao}},\ and\
  \bibinfo {author} {\bibfnamefont {Z.}~\bibnamefont {Chang}},\ }\bibfield
  {title} {\bibinfo {title} {The generation, characterization and applications
  of broadband isolated attosecond pulses},\ }\href
  {http://dx.doi.org/10.1038/nphoton.2013.362} {\bibfield  {journal} {\bibinfo
  {journal} {Nat Photon}\ }\textbf {\bibinfo {volume} {8}},\ \bibinfo {pages}
  {178} (\bibinfo {year} {2014})}\BibitemShut {NoStop}%
\bibitem [{\citenamefont {Krausz}\ and\ \citenamefont
  {Ivanov}(2009)}]{Krausz_RMP2009}%
  \BibitemOpen
  \bibfield  {author} {\bibinfo {author} {\bibfnamefont {F.}~\bibnamefont
  {Krausz}}\ and\ \bibinfo {author} {\bibfnamefont {M.}~\bibnamefont
  {Ivanov}},\ }\bibfield  {title} {\bibinfo {title} {Attosecond physics},\
  }\href {https://doi.org/10.1103/RevModPhys.81.163} {\bibfield  {journal}
  {\bibinfo  {journal} {Rev. Mod. Phys.}\ }\textbf {\bibinfo {volume} {81}},\
  \bibinfo {pages} {163} (\bibinfo {year} {2009})}\BibitemShut {NoStop}%
\bibitem [{\citenamefont {Schn\"urer}\ \emph {et~al.}(1998)\citenamefont
  {Schn\"urer}, \citenamefont {Spielmann}, \citenamefont {Wobrauschek},
  \citenamefont {Streli}, \citenamefont {Burnett}, \citenamefont {Kan},
  \citenamefont {Ferencz}, \citenamefont {Koppitsch}, \citenamefont {Cheng},
  \citenamefont {Brabec},\ and\ \citenamefont {Krausz}}]{Schnurer_PRL1998}%
  \BibitemOpen
  \bibfield  {author} {\bibinfo {author} {\bibfnamefont {M.}~\bibnamefont
  {Schn\"urer}}, \bibinfo {author} {\bibfnamefont {C.}~\bibnamefont
  {Spielmann}}, \bibinfo {author} {\bibfnamefont {P.}~\bibnamefont
  {Wobrauschek}}, \bibinfo {author} {\bibfnamefont {C.}~\bibnamefont {Streli}},
  \bibinfo {author} {\bibfnamefont {N.~H.}\ \bibnamefont {Burnett}}, \bibinfo
  {author} {\bibfnamefont {C.}~\bibnamefont {Kan}}, \bibinfo {author}
  {\bibfnamefont {K.}~\bibnamefont {Ferencz}}, \bibinfo {author} {\bibfnamefont
  {R.}~\bibnamefont {Koppitsch}}, \bibinfo {author} {\bibfnamefont
  {Z.}~\bibnamefont {Cheng}}, \bibinfo {author} {\bibfnamefont
  {T.}~\bibnamefont {Brabec}},\ and\ \bibinfo {author} {\bibfnamefont
  {F.}~\bibnamefont {Krausz}},\ }\bibfield  {title} {\bibinfo {title} {Coherent
  0.5-kev x-ray emission from helium driven by a sub-10-fs laser},\ }\href
  {https://doi.org/10.1103/PhysRevLett.80.3236} {\bibfield  {journal} {\bibinfo
   {journal} {Phys. Rev. Lett.}\ }\textbf {\bibinfo {volume} {80}},\ \bibinfo
  {pages} {3236} (\bibinfo {year} {1998})}\BibitemShut {NoStop}%
\bibitem [{\citenamefont {Popmintchev}\ \emph {et~al.}(2012)\citenamefont
  {Popmintchev}, \citenamefont {Chen}, \citenamefont {Popmintchev},
  \citenamefont {Arpin}, \citenamefont {Brown}, \citenamefont {Ali{\v
  s}auskas}, \citenamefont {Andriukaitis}, \citenamefont {Bal{\v c}iunas},
  \citenamefont {M{\"u}cke}, \citenamefont {Pugzlys}, \citenamefont {Baltu{\v
  s}ka}, \citenamefont {Shim}, \citenamefont {Schrauth}, \citenamefont {Gaeta},
  \citenamefont {Hern{\'a}ndez-Garc{\'\i}a}, \citenamefont {Plaja},
  \citenamefont {Becker}, \citenamefont {Jaron-Becker}, \citenamefont
  {Murnane},\ and\ \citenamefont {Kapteyn}}]{Popmintchev_Science2012}%
  \BibitemOpen
  \bibfield  {author} {\bibinfo {author} {\bibfnamefont {T.}~\bibnamefont
  {Popmintchev}}, \bibinfo {author} {\bibfnamefont {M.-C.}\ \bibnamefont
  {Chen}}, \bibinfo {author} {\bibfnamefont {D.}~\bibnamefont {Popmintchev}},
  \bibinfo {author} {\bibfnamefont {P.}~\bibnamefont {Arpin}}, \bibinfo
  {author} {\bibfnamefont {S.}~\bibnamefont {Brown}}, \bibinfo {author}
  {\bibfnamefont {S.}~\bibnamefont {Ali{\v s}auskas}}, \bibinfo {author}
  {\bibfnamefont {G.}~\bibnamefont {Andriukaitis}}, \bibinfo {author}
  {\bibfnamefont {T.}~\bibnamefont {Bal{\v c}iunas}}, \bibinfo {author}
  {\bibfnamefont {O.~D.}\ \bibnamefont {M{\"u}cke}}, \bibinfo {author}
  {\bibfnamefont {A.}~\bibnamefont {Pugzlys}}, \bibinfo {author} {\bibfnamefont
  {A.}~\bibnamefont {Baltu{\v s}ka}}, \bibinfo {author} {\bibfnamefont
  {B.}~\bibnamefont {Shim}}, \bibinfo {author} {\bibfnamefont {S.~E.}\
  \bibnamefont {Schrauth}}, \bibinfo {author} {\bibfnamefont {A.}~\bibnamefont
  {Gaeta}}, \bibinfo {author} {\bibfnamefont {C.}~\bibnamefont
  {Hern{\'a}ndez-Garc{\'\i}a}}, \bibinfo {author} {\bibfnamefont
  {L.}~\bibnamefont {Plaja}}, \bibinfo {author} {\bibfnamefont
  {A.}~\bibnamefont {Becker}}, \bibinfo {author} {\bibfnamefont
  {A.}~\bibnamefont {Jaron-Becker}}, \bibinfo {author} {\bibfnamefont {M.~M.}\
  \bibnamefont {Murnane}},\ and\ \bibinfo {author} {\bibfnamefont {H.~C.}\
  \bibnamefont {Kapteyn}},\ }\bibfield  {title} {\bibinfo {title} {Bright
  coherent ultrahigh harmonics in the kev x-ray regime from mid-infrared
  femtosecond lasers},\ }\href {https://doi.org/10.1126/science.1218497}
  {\bibfield  {journal} {\bibinfo  {journal} {Science}\ }\textbf {\bibinfo
  {volume} {336}},\ \bibinfo {pages} {1287} (\bibinfo {year} {2012})},\ \Eprint
  {https://arxiv.org/abs/http://www.sciencemag.org/content/336/6086/1287.full.pdf}
  {http://www.sciencemag.org/content/336/6086/1287.full.pdf} \BibitemShut
  {NoStop}%
\bibitem [{\citenamefont {Boyd}(2003)}]{Boyd:2003}%
  \BibitemOpen
  \bibfield  {author} {\bibinfo {author} {\bibfnamefont {R.}~\bibnamefont
  {Boyd}},\ }\href@noop {} {\emph {\bibinfo {title} {Nonlinear Optics}}},\
  \bibinfo {edition} {2nd}\ ed.\ (\bibinfo  {publisher} {Academic Press},\
  \bibinfo {address} {Amsterdam},\ \bibinfo {year} {2003})\BibitemShut
  {NoStop}%
\bibitem [{\citenamefont {DeSalvo}\ \emph {et~al.}(1992)\citenamefont
  {DeSalvo}, \citenamefont {Hagan}, \citenamefont {Sheik-Bahae}, \citenamefont
  {Stegeman}, \citenamefont {Stryland},\ and\ \citenamefont
  {Vanherzeele}}]{DeSalvo_OptLett1992}%
  \BibitemOpen
  \bibfield  {author} {\bibinfo {author} {\bibfnamefont {R.}~\bibnamefont
  {DeSalvo}}, \bibinfo {author} {\bibfnamefont {D.~J.}\ \bibnamefont {Hagan}},
  \bibinfo {author} {\bibfnamefont {M.}~\bibnamefont {Sheik-Bahae}}, \bibinfo
  {author} {\bibfnamefont {G.}~\bibnamefont {Stegeman}}, \bibinfo {author}
  {\bibfnamefont {E.~W.~V.}\ \bibnamefont {Stryland}},\ and\ \bibinfo {author}
  {\bibfnamefont {H.}~\bibnamefont {Vanherzeele}},\ }\bibfield  {title}
  {\bibinfo {title} {Self-focusing and self-defocusing by cascaded second-order
  effects in ktp},\ }\href {https://doi.org/10.1364/OL.17.000028} {\bibfield
  {journal} {\bibinfo  {journal} {Opt. Lett.}\ }\textbf {\bibinfo {volume}
  {17}},\ \bibinfo {pages} {28} (\bibinfo {year} {1992})}\BibitemShut {NoStop}%
\bibitem [{\citenamefont {Liu}\ \emph {et~al.}(1999)\citenamefont {Liu},
  \citenamefont {Qian},\ and\ \citenamefont {Wise}}]{Liu_OptLett1999}%
  \BibitemOpen
  \bibfield  {author} {\bibinfo {author} {\bibfnamefont {X.}~\bibnamefont
  {Liu}}, \bibinfo {author} {\bibfnamefont {L.}~\bibnamefont {Qian}},\ and\
  \bibinfo {author} {\bibfnamefont {F.}~\bibnamefont {Wise}},\ }\bibfield
  {title} {\bibinfo {title} {High-energy pulse compression by use of negative
  phase shifts produced by the cascade $\chi^{(2)}:\chi^{(2)}$ nonlinearity},\
  }\href {https://doi.org/10.1364/OL.24.001777} {\bibfield  {journal} {\bibinfo
   {journal} {Opt. Lett.}\ }\textbf {\bibinfo {volume} {24}},\ \bibinfo {pages}
  {1777} (\bibinfo {year} {1999})}\BibitemShut {NoStop}%
\bibitem [{\citenamefont {Couch}\ \emph {et~al.}(2020)\citenamefont {Couch},
  \citenamefont {Hickstein}, \citenamefont {Winters}, \citenamefont {Backus},
  \citenamefont {Kirchner}, \citenamefont {Domingue}, \citenamefont {Ramirez},
  \citenamefont {Durfee}, \citenamefont {Murnane},\ and\ \citenamefont
  {Kapteyn}}]{Couch_Optica2020}%
  \BibitemOpen
  \bibfield  {author} {\bibinfo {author} {\bibfnamefont {D.~E.}\ \bibnamefont
  {Couch}}, \bibinfo {author} {\bibfnamefont {D.~D.}\ \bibnamefont
  {Hickstein}}, \bibinfo {author} {\bibfnamefont {D.~G.}\ \bibnamefont
  {Winters}}, \bibinfo {author} {\bibfnamefont {S.~J.}\ \bibnamefont {Backus}},
  \bibinfo {author} {\bibfnamefont {M.~S.}\ \bibnamefont {Kirchner}}, \bibinfo
  {author} {\bibfnamefont {S.~R.}\ \bibnamefont {Domingue}}, \bibinfo {author}
  {\bibfnamefont {J.~J.}\ \bibnamefont {Ramirez}}, \bibinfo {author}
  {\bibfnamefont {C.~G.}\ \bibnamefont {Durfee}}, \bibinfo {author}
  {\bibfnamefont {M.~M.}\ \bibnamefont {Murnane}},\ and\ \bibinfo {author}
  {\bibfnamefont {H.~C.}\ \bibnamefont {Kapteyn}},\ }\bibfield  {title}
  {\bibinfo {title} {Ultrafast 1 mhz vacuum-ultraviolet source via highly
  cascaded harmonic generation in negative-curvature hollow-core fibers},\
  }\href@noop {} {\bibfield  {journal} {\bibinfo  {journal} {Optica}\ }\textbf
  {\bibinfo {volume} {7}},\ \bibinfo {pages} {832} (\bibinfo {year}
  {2020})}\BibitemShut {NoStop}%
\bibitem [{\citenamefont {Ghimire}\ and\ \citenamefont
  {Reis}(2019)}]{Ghimire_NatPhys2019}%
  \BibitemOpen
  \bibfield  {author} {\bibinfo {author} {\bibfnamefont {S.}~\bibnamefont
  {Ghimire}}\ and\ \bibinfo {author} {\bibfnamefont {D.~A.}\ \bibnamefont
  {Reis}},\ }\bibfield  {title} {\bibinfo {title} {High-harmonic generation
  from solids},\ }\href {https://doi.org/10.1038/s41567-018-0315-5} {\bibfield
  {journal} {\bibinfo  {journal} {Nature Physics}\ }\textbf {\bibinfo {volume}
  {15}},\ \bibinfo {pages} {10} (\bibinfo {year} {2019})}\BibitemShut {NoStop}%
\bibitem [{\citenamefont {Liu}\ \emph {et~al.}(2016{\natexlab{a}})\citenamefont
  {Liu}, \citenamefont {Li}, \citenamefont {You}, \citenamefont {Ghimire},
  \citenamefont {Heinz},\ and\ \citenamefont {Reis}}]{Liu_NatPhys2017}%
  \BibitemOpen
  \bibfield  {author} {\bibinfo {author} {\bibfnamefont {H.}~\bibnamefont
  {Liu}}, \bibinfo {author} {\bibfnamefont {Y.}~\bibnamefont {Li}}, \bibinfo
  {author} {\bibfnamefont {Y.~S.}\ \bibnamefont {You}}, \bibinfo {author}
  {\bibfnamefont {S.}~\bibnamefont {Ghimire}}, \bibinfo {author} {\bibfnamefont
  {T.~F.}\ \bibnamefont {Heinz}},\ and\ \bibinfo {author} {\bibfnamefont
  {D.~A.}\ \bibnamefont {Reis}},\ }\bibfield  {title} {\bibinfo {title}
  {High-harmonic generation from an atomically thin semiconductor},\ }\href
  {https://doi.org/10.1038/nphys3946} {\bibfield  {journal} {\bibinfo
  {journal} {Nature Physics}\ }\textbf {\bibinfo {volume} {13}},\ \bibinfo
  {pages} {262 EP } (\bibinfo {year} {2016}{\natexlab{a}})}\BibitemShut
  {NoStop}%
\bibitem [{\citenamefont {Liu}\ \emph {et~al.}(2018)\citenamefont {Liu},
  \citenamefont {Guo}, \citenamefont {Vampa}, \citenamefont {Zhang},
  \citenamefont {Sarmiento}, \citenamefont {Xiao}, \citenamefont {Bucksbaum},
  \citenamefont {Vu{\v c}kovi{\'c}}, \citenamefont {Fan},\ and\ \citenamefont
  {Reis}}]{Liu_NatPhys2018}%
  \BibitemOpen
  \bibfield  {author} {\bibinfo {author} {\bibfnamefont {H.}~\bibnamefont
  {Liu}}, \bibinfo {author} {\bibfnamefont {C.}~\bibnamefont {Guo}}, \bibinfo
  {author} {\bibfnamefont {G.}~\bibnamefont {Vampa}}, \bibinfo {author}
  {\bibfnamefont {J.~L.}\ \bibnamefont {Zhang}}, \bibinfo {author}
  {\bibfnamefont {T.}~\bibnamefont {Sarmiento}}, \bibinfo {author}
  {\bibfnamefont {M.}~\bibnamefont {Xiao}}, \bibinfo {author} {\bibfnamefont
  {P.~H.}\ \bibnamefont {Bucksbaum}}, \bibinfo {author} {\bibfnamefont
  {J.}~\bibnamefont {Vu{\v c}kovi{\'c}}}, \bibinfo {author} {\bibfnamefont
  {S.}~\bibnamefont {Fan}},\ and\ \bibinfo {author} {\bibfnamefont {D.~A.}\
  \bibnamefont {Reis}},\ }\bibfield  {title} {\bibinfo {title} {Enhanced
  high-harmonic generation from an all-dielectric metasurface},\ }\href
  {https://doi.org/10.1038/s41567-018-0233-6} {\bibfield  {journal} {\bibinfo
  {journal} {Nature Physics}\ }\textbf {\bibinfo {volume} {14}},\ \bibinfo
  {pages} {1006} (\bibinfo {year} {2018})}\BibitemShut {NoStop}%
\bibitem [{\citenamefont {Park}\ \emph {et~al.}(2017)\citenamefont {Park},
  \citenamefont {Camper}, \citenamefont {Kafka}, \citenamefont {Ma},
  \citenamefont {Lai}, \citenamefont {Blaga}, \citenamefont {Agostini},
  \citenamefont {DiMauro},\ and\ \citenamefont {Chowdhury}}]{Park_OptLett2017}%
  \BibitemOpen
  \bibfield  {author} {\bibinfo {author} {\bibfnamefont {H.}~\bibnamefont
  {Park}}, \bibinfo {author} {\bibfnamefont {A.}~\bibnamefont {Camper}},
  \bibinfo {author} {\bibfnamefont {K.}~\bibnamefont {Kafka}}, \bibinfo
  {author} {\bibfnamefont {B.}~\bibnamefont {Ma}}, \bibinfo {author}
  {\bibfnamefont {Y.~H.}\ \bibnamefont {Lai}}, \bibinfo {author} {\bibfnamefont
  {C.}~\bibnamefont {Blaga}}, \bibinfo {author} {\bibfnamefont
  {P.}~\bibnamefont {Agostini}}, \bibinfo {author} {\bibfnamefont {L.~F.}\
  \bibnamefont {DiMauro}},\ and\ \bibinfo {author} {\bibfnamefont
  {E.}~\bibnamefont {Chowdhury}},\ }\bibfield  {title} {\bibinfo {title}
  {High-order harmonic generations in intense mir fields by cascade three-wave
  mixing in a fractal-poled linbo3 photonic crystal},\ }\href
  {https://doi.org/10.1364/OL.42.004020} {\bibfield  {journal} {\bibinfo
  {journal} {Opt. Lett.}\ }\textbf {\bibinfo {volume} {42}},\ \bibinfo {pages}
  {4020} (\bibinfo {year} {2017})}\BibitemShut {NoStop}%
\bibitem [{\citenamefont {Hickstein}\ \emph {et~al.}(2017)\citenamefont
  {Hickstein}, \citenamefont {Carlson}, \citenamefont {Kowligy}, \citenamefont
  {Kirchner}, \citenamefont {Domingue}, \citenamefont {Nader}, \citenamefont
  {Timmers}, \citenamefont {Lind}, \citenamefont {Ycas}, \citenamefont
  {Murnane}, \citenamefont {Kapteyn}, \citenamefont {Papp},\ and\ \citenamefont
  {Diddams}}]{Hickstein_Optica2017}%
  \BibitemOpen
  \bibfield  {author} {\bibinfo {author} {\bibfnamefont {D.~D.}\ \bibnamefont
  {Hickstein}}, \bibinfo {author} {\bibfnamefont {D.~R.}\ \bibnamefont
  {Carlson}}, \bibinfo {author} {\bibfnamefont {A.}~\bibnamefont {Kowligy}},
  \bibinfo {author} {\bibfnamefont {M.}~\bibnamefont {Kirchner}}, \bibinfo
  {author} {\bibfnamefont {S.~R.}\ \bibnamefont {Domingue}}, \bibinfo {author}
  {\bibfnamefont {N.}~\bibnamefont {Nader}}, \bibinfo {author} {\bibfnamefont
  {H.}~\bibnamefont {Timmers}}, \bibinfo {author} {\bibfnamefont
  {A.}~\bibnamefont {Lind}}, \bibinfo {author} {\bibfnamefont {G.~G.}\
  \bibnamefont {Ycas}}, \bibinfo {author} {\bibfnamefont {M.~M.}\ \bibnamefont
  {Murnane}}, \bibinfo {author} {\bibfnamefont {H.~C.}\ \bibnamefont
  {Kapteyn}}, \bibinfo {author} {\bibfnamefont {S.~B.}\ \bibnamefont {Papp}},\
  and\ \bibinfo {author} {\bibfnamefont {S.~A.}\ \bibnamefont {Diddams}},\
  }\bibfield  {title} {\bibinfo {title} {High-harmonic generation in
  periodically poled waveguides},\ }\href
  {https://doi.org/10.1364/OPTICA.4.001538} {\bibfield  {journal} {\bibinfo
  {journal} {Optica}\ }\textbf {\bibinfo {volume} {4}},\ \bibinfo {pages}
  {1538} (\bibinfo {year} {2017})}\BibitemShut {NoStop}%
\bibitem [{\citenamefont {Hickstein}\ \emph {et~al.}(2018)\citenamefont
  {Hickstein}, \citenamefont {Carlson}, \citenamefont {Kowligy}, \citenamefont
  {Domingue}, \citenamefont {Kirchner}, \citenamefont {Timmers}, \citenamefont
  {Nader}, \citenamefont {Lind}, \citenamefont {Guo}, \citenamefont
  {Herkommer}, \citenamefont {Kippenberg}, \citenamefont {Murnane},
  \citenamefont {Kapteyn}, \citenamefont {Papp},\ and\ \citenamefont
  {Diddams}}]{Hickstein_CLEO2018}%
  \BibitemOpen
  \bibfield  {author} {\bibinfo {author} {\bibfnamefont {D.~D.}\ \bibnamefont
  {Hickstein}}, \bibinfo {author} {\bibfnamefont {D.~R.}\ \bibnamefont
  {Carlson}}, \bibinfo {author} {\bibfnamefont {A.}~\bibnamefont {Kowligy}},
  \bibinfo {author} {\bibfnamefont {S.~R.}\ \bibnamefont {Domingue}}, \bibinfo
  {author} {\bibfnamefont {M.}~\bibnamefont {Kirchner}}, \bibinfo {author}
  {\bibfnamefont {H.}~\bibnamefont {Timmers}}, \bibinfo {author} {\bibfnamefont
  {N.}~\bibnamefont {Nader}}, \bibinfo {author} {\bibfnamefont
  {A.}~\bibnamefont {Lind}}, \bibinfo {author} {\bibfnamefont {H.}~\bibnamefont
  {Guo}}, \bibinfo {author} {\bibfnamefont {C.}~\bibnamefont {Herkommer}},
  \bibinfo {author} {\bibfnamefont {T.}~\bibnamefont {Kippenberg}}, \bibinfo
  {author} {\bibfnamefont {M.~M.}\ \bibnamefont {Murnane}}, \bibinfo {author}
  {\bibfnamefont {H.~C.}\ \bibnamefont {Kapteyn}}, \bibinfo {author}
  {\bibfnamefont {S.~B.}\ \bibnamefont {Papp}},\ and\ \bibinfo {author}
  {\bibfnamefont {S.~A.}\ \bibnamefont {Diddams}},\ }\bibfield  {title}
  {\bibinfo {title} {Nanophotonic waveguides for extreme nonlinear optics},\
  }in\ \href {https://doi.org/10.1364/CLEO_QELS.2018.FF2E.4} {\emph {\bibinfo
  {booktitle} {Conference on Lasers and Electro-Optics}}}\ (\bibinfo
  {publisher} {Optical Society of America},\ \bibinfo {year} {2018})\ p.\
  \bibinfo {pages} {FF2E.4}\BibitemShut {NoStop}%
\bibitem [{\citenamefont {Dudley}\ \emph {et~al.}(2006)\citenamefont {Dudley},
  \citenamefont {Genty},\ and\ \citenamefont {Coen}}]{Dudley_2006}%
  \BibitemOpen
  \bibfield  {author} {\bibinfo {author} {\bibfnamefont {J.~M.}\ \bibnamefont
  {Dudley}}, \bibinfo {author} {\bibfnamefont {G.}~\bibnamefont {Genty}},\ and\
  \bibinfo {author} {\bibfnamefont {S.}~\bibnamefont {Coen}},\ }\bibfield
  {title} {\bibinfo {title} {Supercontinuum generation in photonic crystal
  fiber},\ }\href {https://doi.org/10.1103/RevModPhys.78.1135} {\bibfield
  {journal} {\bibinfo  {journal} {Rev. Mod. Phys.}\ }\textbf {\bibinfo {volume}
  {78}},\ \bibinfo {pages} {1135} (\bibinfo {year} {2006})}\BibitemShut
  {NoStop}%
\bibitem [{\citenamefont {Catanese}\ \emph {et~al.}(2020)\citenamefont
  {Catanese}, \citenamefont {Rutledge}, \citenamefont {Silfies}, \citenamefont
  {Li}, \citenamefont {Timmers}, \citenamefont {Kowligy}, \citenamefont {Lind},
  \citenamefont {Diddams},\ and\ \citenamefont
  {Allison}}]{Catanese_OptLett2020}%
  \BibitemOpen
  \bibfield  {author} {\bibinfo {author} {\bibfnamefont {A.}~\bibnamefont
  {Catanese}}, \bibinfo {author} {\bibfnamefont {J.}~\bibnamefont {Rutledge}},
  \bibinfo {author} {\bibfnamefont {M.~C.}\ \bibnamefont {Silfies}}, \bibinfo
  {author} {\bibfnamefont {X.}~\bibnamefont {Li}}, \bibinfo {author}
  {\bibfnamefont {H.}~\bibnamefont {Timmers}}, \bibinfo {author} {\bibfnamefont
  {A.~S.}\ \bibnamefont {Kowligy}}, \bibinfo {author} {\bibfnamefont
  {A.}~\bibnamefont {Lind}}, \bibinfo {author} {\bibfnamefont {S.~A.}\
  \bibnamefont {Diddams}},\ and\ \bibinfo {author} {\bibfnamefont {T.~K.}\
  \bibnamefont {Allison}},\ }\bibfield  {title} {\bibinfo {title} {Mid-infrared
  frequency comb with 6.7 w average power based on difference frequency
  generation},\ }\href {https://doi.org/10.1364/OL.385294} {\bibfield
  {journal} {\bibinfo  {journal} {Opt. Lett.}\ }\textbf {\bibinfo {volume}
  {45}},\ \bibinfo {pages} {1248} (\bibinfo {year} {2020})}\BibitemShut
  {NoStop}%
\bibitem [{\citenamefont {Conforti}\ \emph {et~al.}(2010)\citenamefont
  {Conforti}, \citenamefont {Baronio},\ and\ \citenamefont
  {De~Angelis}}]{Conforti_PRA2010}%
  \BibitemOpen
  \bibfield  {author} {\bibinfo {author} {\bibfnamefont {M.}~\bibnamefont
  {Conforti}}, \bibinfo {author} {\bibfnamefont {F.}~\bibnamefont {Baronio}},\
  and\ \bibinfo {author} {\bibfnamefont {C.}~\bibnamefont {De~Angelis}},\
  }\bibfield  {title} {\bibinfo {title} {Nonlinear envelope equation for
  broadband optical pulses in quadratic media},\ }\href
  {https://doi.org/10.1103/PhysRevA.81.053841} {\bibfield  {journal} {\bibinfo
  {journal} {Phys. Rev. A}\ }\textbf {\bibinfo {volume} {81}},\ \bibinfo
  {pages} {053841} (\bibinfo {year} {2010})}\BibitemShut {NoStop}%
\bibitem [{\citenamefont {{Conforti}}\ \emph {et~al.}(2010)\citenamefont
  {{Conforti}}, \citenamefont {{Baronio}},\ and\ \citenamefont {{De
  Angelis}}}]{Conforti_IEEE2010}%
  \BibitemOpen
  \bibfield  {author} {\bibinfo {author} {\bibfnamefont {M.}~\bibnamefont
  {{Conforti}}}, \bibinfo {author} {\bibfnamefont {F.}~\bibnamefont
  {{Baronio}}},\ and\ \bibinfo {author} {\bibfnamefont {C.}~\bibnamefont {{De
  Angelis}}},\ }\bibfield  {title} {\bibinfo {title} {Ultrabroadband optical
  phenomena in quadratic nonlinear media},\ }\href@noop {} {\bibfield
  {journal} {\bibinfo  {journal} {IEEE Photonics Journal}\ }\textbf {\bibinfo
  {volume} {2}},\ \bibinfo {pages} {600} (\bibinfo {year} {2010})}\BibitemShut
  {NoStop}%
\bibitem [{\citenamefont {Castillo-Torres}(2013)}]{Castill-Torres_OptComm2013}%
  \BibitemOpen
  \bibfield  {author} {\bibinfo {author} {\bibfnamefont {J.}~\bibnamefont
  {Castillo-Torres}},\ }\bibfield  {title} {\bibinfo {title} {Optical
  absorption edge analysis for zinc-doped lithium niobate},\ }\href
  {https://doi.org/https://doi.org/10.1016/j.optcom.2012.10.067} {\bibfield
  {journal} {\bibinfo  {journal} {Optics Communications}\ }\textbf {\bibinfo
  {volume} {290}},\ \bibinfo {pages} {107 } (\bibinfo {year}
  {2013})}\BibitemShut {NoStop}%
\bibitem [{\citenamefont {Kowligy}\ \emph {et~al.}(2018)\citenamefont
  {Kowligy}, \citenamefont {Lind}, \citenamefont {Hickstein}, \citenamefont
  {Carlson}, \citenamefont {Timmers}, \citenamefont {Nader}, \citenamefont
  {Cruz}, \citenamefont {Ycas}, \citenamefont {Papp},\ and\ \citenamefont
  {Diddams}}]{Kowligy_OptLett2018}%
  \BibitemOpen
  \bibfield  {author} {\bibinfo {author} {\bibfnamefont {A.~S.}\ \bibnamefont
  {Kowligy}}, \bibinfo {author} {\bibfnamefont {A.}~\bibnamefont {Lind}},
  \bibinfo {author} {\bibfnamefont {D.~D.}\ \bibnamefont {Hickstein}}, \bibinfo
  {author} {\bibfnamefont {D.~R.}\ \bibnamefont {Carlson}}, \bibinfo {author}
  {\bibfnamefont {H.}~\bibnamefont {Timmers}}, \bibinfo {author} {\bibfnamefont
  {N.}~\bibnamefont {Nader}}, \bibinfo {author} {\bibfnamefont {F.~C.}\
  \bibnamefont {Cruz}}, \bibinfo {author} {\bibfnamefont {G.}~\bibnamefont
  {Ycas}}, \bibinfo {author} {\bibfnamefont {S.~B.}\ \bibnamefont {Papp}},\
  and\ \bibinfo {author} {\bibfnamefont {S.~A.}\ \bibnamefont {Diddams}},\
  }\bibfield  {title} {\bibinfo {title} {Mid-infrared frequency comb generation
  via cascaded quadratic nonlinearities in quasi-phase-matched waveguides},\
  }\href {https://doi.org/10.1364/OL.43.001678} {\bibfield  {journal} {\bibinfo
   {journal} {Opt. Lett.}\ }\textbf {\bibinfo {volume} {43}},\ \bibinfo {pages}
  {1678} (\bibinfo {year} {2018})}\BibitemShut {NoStop}%
\bibitem [{\citenamefont {Phillips}\ \emph
  {et~al.}(2011{\natexlab{a}})\citenamefont {Phillips}, \citenamefont
  {Langrock}, \citenamefont {Pelc}, \citenamefont {Fejer}, \citenamefont
  {Hartl},\ and\ \citenamefont {Fermann}}]{Phillips_OptExp2011}%
  \BibitemOpen
  \bibfield  {author} {\bibinfo {author} {\bibfnamefont {C.~R.}\ \bibnamefont
  {Phillips}}, \bibinfo {author} {\bibfnamefont {C.}~\bibnamefont {Langrock}},
  \bibinfo {author} {\bibfnamefont {J.~S.}\ \bibnamefont {Pelc}}, \bibinfo
  {author} {\bibfnamefont {M.~M.}\ \bibnamefont {Fejer}}, \bibinfo {author}
  {\bibfnamefont {I.}~\bibnamefont {Hartl}},\ and\ \bibinfo {author}
  {\bibfnamefont {M.~E.}\ \bibnamefont {Fermann}},\ }\bibfield  {title}
  {\bibinfo {title} {Supercontinuum generation in quasi-phasematched
  waveguides},\ }\href {https://doi.org/10.1364/OE.19.018754} {\bibfield
  {journal} {\bibinfo  {journal} {Opt. Express}\ }\textbf {\bibinfo {volume}
  {19}},\ \bibinfo {pages} {18754} (\bibinfo {year}
  {2011}{\natexlab{a}})}\BibitemShut {NoStop}%
\bibitem [{\citenamefont {Lind}\ \emph {et~al.}(2020)\citenamefont {Lind},
  \citenamefont {Kowligy}, \citenamefont {Timmers}, \citenamefont {Cruz},
  \citenamefont {Nader}, \citenamefont {Silfies}, \citenamefont {Allison},\
  and\ \citenamefont {Diddams}}]{Lind_PRL2020}%
  \BibitemOpen
  \bibfield  {author} {\bibinfo {author} {\bibfnamefont {A.~J.}\ \bibnamefont
  {Lind}}, \bibinfo {author} {\bibfnamefont {A.}~\bibnamefont {Kowligy}},
  \bibinfo {author} {\bibfnamefont {H.}~\bibnamefont {Timmers}}, \bibinfo
  {author} {\bibfnamefont {F.~C.}\ \bibnamefont {Cruz}}, \bibinfo {author}
  {\bibfnamefont {N.}~\bibnamefont {Nader}}, \bibinfo {author} {\bibfnamefont
  {M.~C.}\ \bibnamefont {Silfies}}, \bibinfo {author} {\bibfnamefont {T.~K.}\
  \bibnamefont {Allison}},\ and\ \bibinfo {author} {\bibfnamefont {S.~A.}\
  \bibnamefont {Diddams}},\ }\bibfield  {title} {\bibinfo {title} {Mid-infrared
  frequency comb generation and spectroscopy with few-cycle pulses and
  ${\ensuremath{\chi}}^{(2)}$ nonlinear optics},\ }\href
  {https://doi.org/10.1103/PhysRevLett.124.133904} {\bibfield  {journal}
  {\bibinfo  {journal} {Phys. Rev. Lett.}\ }\textbf {\bibinfo {volume} {124}},\
  \bibinfo {pages} {133904} (\bibinfo {year} {2020})}\BibitemShut {NoStop}%
\bibitem [{\citenamefont {Agrawal}(2012)}]{Agrawal_NonlinearFiberOpticsBook}%
  \BibitemOpen
  \bibfield  {author} {\bibinfo {author} {\bibfnamefont {G.~P.}\ \bibnamefont
  {Agrawal}},\ }\href@noop {} {\emph {\bibinfo {title} {Nonlinear Fiber
  Optics}}},\ edited by\ \bibinfo {editor} {\bibnamefont {5}}\ (\bibinfo
  {publisher} {Academic Press},\ \bibinfo {year} {2012})\BibitemShut {NoStop}%
\bibitem [{COM()}]{COMSOL_Ref}%
  \BibitemOpen
  \href@noop {} {\bibinfo {title} {Comsol multiphysics{\textregistered} v. 5.4.
  www.comsol.com. comsol ab, stockholm, sweden.}}\BibitemShut {Stop}%
\bibitem [{\citenamefont {Shoji}\ \emph {et~al.}(1997)\citenamefont {Shoji},
  \citenamefont {Kondo}, \citenamefont {Kitamoto}, \citenamefont {Shirane},\
  and\ \citenamefont {Ito}}]{Shoji_JOSAB1997}%
  \BibitemOpen
  \bibfield  {author} {\bibinfo {author} {\bibfnamefont {I.}~\bibnamefont
  {Shoji}}, \bibinfo {author} {\bibfnamefont {T.}~\bibnamefont {Kondo}},
  \bibinfo {author} {\bibfnamefont {A.}~\bibnamefont {Kitamoto}}, \bibinfo
  {author} {\bibfnamefont {M.}~\bibnamefont {Shirane}},\ and\ \bibinfo {author}
  {\bibfnamefont {R.}~\bibnamefont {Ito}},\ }\bibfield  {title} {\bibinfo
  {title} {Absolute scale of second-order nonlinear-optical coefficients},\
  }\href {https://doi.org/10.1364/JOSAB.14.002268} {\bibfield  {journal}
  {\bibinfo  {journal} {J. Opt. Soc. Am. B}\ }\textbf {\bibinfo {volume}
  {14}},\ \bibinfo {pages} {2268} (\bibinfo {year} {1997})}\BibitemShut
  {NoStop}%
\bibitem [{\citenamefont {Balac}\ and\ \citenamefont
  {Mah{\'e}}(2013)}]{Balac_CompPhysComm2013}%
  \BibitemOpen
  \bibfield  {author} {\bibinfo {author} {\bibfnamefont {S.}~\bibnamefont
  {Balac}}\ and\ \bibinfo {author} {\bibfnamefont {F.}~\bibnamefont
  {Mah{\'e}}},\ }\bibfield  {title} {\bibinfo {title} {Embedded runge--kutta
  scheme for step-size control in the interaction picture method},\ }\href
  {https://doi.org/https://doi.org/10.1016/j.cpc.2012.12.020} {\bibfield
  {journal} {\bibinfo  {journal} {Computer Physics Communications}\ }\textbf
  {\bibinfo {volume} {184}},\ \bibinfo {pages} {1211} (\bibinfo {year}
  {2013})}\BibitemShut {NoStop}%
\bibitem [{\citenamefont {Phillips}\ \emph
  {et~al.}(2011{\natexlab{b}})\citenamefont {Phillips}, \citenamefont
  {Langrock}, \citenamefont {Pelc}, \citenamefont {Fejer}, \citenamefont
  {Jiang}, \citenamefont {Fermann},\ and\ \citenamefont
  {Hartl}}]{Phillips_OptLett2011_2}%
  \BibitemOpen
  \bibfield  {author} {\bibinfo {author} {\bibfnamefont {C.~R.}\ \bibnamefont
  {Phillips}}, \bibinfo {author} {\bibfnamefont {C.}~\bibnamefont {Langrock}},
  \bibinfo {author} {\bibfnamefont {J.~S.}\ \bibnamefont {Pelc}}, \bibinfo
  {author} {\bibfnamefont {M.~M.}\ \bibnamefont {Fejer}}, \bibinfo {author}
  {\bibfnamefont {J.}~\bibnamefont {Jiang}}, \bibinfo {author} {\bibfnamefont
  {M.~E.}\ \bibnamefont {Fermann}},\ and\ \bibinfo {author} {\bibfnamefont
  {I.}~\bibnamefont {Hartl}},\ }\bibfield  {title} {\bibinfo {title}
  {Supercontinuum generation in quasi-phase-matched linbo3 waveguide pumped by
  a tm-doped fiber laser system},\ }\href
  {https://doi.org/10.1364/OL.36.003912} {\bibfield  {journal} {\bibinfo
  {journal} {Opt. Lett.}\ }\textbf {\bibinfo {volume} {36}},\ \bibinfo {pages}
  {3912} (\bibinfo {year} {2011}{\natexlab{b}})}\BibitemShut {NoStop}%
\bibitem [{\citenamefont {Liu}\ \emph {et~al.}(2016{\natexlab{b}})\citenamefont
  {Liu}, \citenamefont {Wright}, \citenamefont {Christodoulides},\ and\
  \citenamefont {Wise}}]{Liu_OptLett2016}%
  \BibitemOpen
  \bibfield  {author} {\bibinfo {author} {\bibfnamefont {Z.}~\bibnamefont
  {Liu}}, \bibinfo {author} {\bibfnamefont {L.~G.}\ \bibnamefont {Wright}},
  \bibinfo {author} {\bibfnamefont {D.~N.}\ \bibnamefont {Christodoulides}},\
  and\ \bibinfo {author} {\bibfnamefont {F.~W.}\ \bibnamefont {Wise}},\
  }\bibfield  {title} {\bibinfo {title} {Kerr self-cleaning of
  femtosecond-pulsed beams in graded-index multimode fiber},\ }\href
  {https://doi.org/10.1364/OL.41.003675} {\bibfield  {journal} {\bibinfo
  {journal} {Opt. Lett.}\ }\textbf {\bibinfo {volume} {41}},\ \bibinfo {pages}
  {3675} (\bibinfo {year} {2016}{\natexlab{b}})}\BibitemShut {NoStop}%
\bibitem [{\citenamefont {Dorozhkin}\ \emph {et~al.}(1976)\citenamefont
  {Dorozhkin}, \citenamefont {Kizel}, \citenamefont {Shigorin},\ and\
  \citenamefont {Shipulo}}]{dorozhkin_dispersion_1976}%
  \BibitemOpen
  \bibfield  {author} {\bibinfo {author} {\bibfnamefont {L.~M.}\ \bibnamefont
  {Dorozhkin}}, \bibinfo {author} {\bibfnamefont {V.~A.}\ \bibnamefont
  {Kizel}}, \bibinfo {author} {\bibfnamefont {V.~D.}\ \bibnamefont
  {Shigorin}},\ and\ \bibinfo {author} {\bibfnamefont {G.~N.}\ \bibnamefont
  {Shipulo}},\ }\bibfield  {title} {\bibinfo {title} {Dispersion of quadratic
  optical susceptibility of lithium niobate and barium sodium niobate
  crystals},\ }\href {http://adsabs.harvard.edu/abs/1976JETPL..24..329D}
  {\bibfield  {journal} {\bibinfo  {journal} {Soviet Journal of Experimental
  and Theoretical Physics Letters}\ }\textbf {\bibinfo {volume} {24}},\
  \bibinfo {pages} {329} (\bibinfo {year} {1976})}\BibitemShut {NoStop}%
\end{thebibliography}


%

\end{document}